\documentclass[acmlarge, nonacm]{acmart}
\AtBeginDocument{%
  }

\usepackage{graphicx}      
\usepackage{bbm}

\newtheorem{mydef}{\bf Definition}

\newtheorem{myprob}{\bf Problem}

\newtheorem{mypro}{\bf Proposition}
\newtheorem{myexm}{\bf Example}
\newtheorem{remark}{\bf Remark}
\newtheorem{assumption}{\bf Assumption}
\usepackage{amsmath}
\usepackage{amsfonts}
\DeclareMathSymbol{\shortminus}{\mathbin}{AMSa}{"39}

\usepackage{enumitem}
\usepackage{multirow}
\usepackage{booktabs}
\usepackage{array}
\usepackage{float}
\usepackage{mathtools}
\usepackage{epsfig}
\usepackage{arydshln}
\usepackage{algpseudocode}
\usepackage{verbatim}
\usepackage{subfigure}
\usepackage{enumerate}
\usepackage{rotating}
\usepackage{color}
\usepackage{flushend}
\usepackage{mathrsfs}
\usepackage[ruled,vlined,linesnumbered]{algorithm2e}
\usepackage{float}
\usepackage{soul}
\usepackage{fancyhdr}
\usepackage{url}
\usepackage{makecell} 
\usepackage[para,online,flushleft]{threeparttable}
\usepackage{tikz}
\usepackage{caption} 
\usepackage{pifont}
\allowdisplaybreaks

\newcommand{\somewhere}{
    \mathbin{\vcenter{\hbox{\begin{tikzpicture}
        \draw (0, 0.14) -- (0.14, 0) -- (0, -0.14) -- (-0.14, 0) -- cycle;
        \draw (0, 0.08) -- (0.08, 0) -- (0, -0.08) -- (-0.08, 0) -- cycle;
    \end{tikzpicture}}}}
}
\newcommand{\everywhere}{
    \mathbin{\vcenter{\hbox{\begin{tikzpicture}
        \draw (-0.1, 0.1) -- (0.1, 0.1) -- (0.1, -0.1) -- (-0.1, -0.1) -- cycle;
        \draw (-0.05, 0.05) -- (0.05, 0.05) -- (0.05, -0.05) -- (-0.05, -0.05) -- cycle;
    \end{tikzpicture}}}}
}

\begin{document}

\title{STL-GO: Spatio-Temporal Logic with Graph Operators for Distributed Systems with Multiple Network Topologies}

\author{Yiqi Zhao}
\authornote{These authors contributed equally to this work.}
\orcid{0009-0007-4283-6358}
\email{yiqizhao@usc.edu}
\affiliation{%
  \institution{University of Southern California}
  \city{Los Angeles}
  \state{California}
  \country{USA}
}

\author{Xinyi Yu}
\orcid{0009-0004-3161-7043}
\authornotemark[1]
\email{xinyi.yu12@usc.edu}
\affiliation{%
  \institution{University of Southern California}
  \city{Los Angeles}
  \state{California}
  \country{USA}
}

\author{Bardh Hoxha}
\orcid{0000-0001-6255-7566}
\email{bardh.hoxha@toyota.com}
\affiliation{%
  \institution{Toyota NA R\&D}
  \city{Ann Arbor}
  \state{Michigan}
  \country{USA}
}

\author{Georgios Fainekos}
\orcid{0000-0002-0456-2129}
\email{georgios.fainekos@toyota.com}
\affiliation{%
  \institution{Toyota NA R\&D}
  \city{Ann Arbor}
  \state{Michigan}
  \country{USA}
}

\author{Jyotirmoy V. Deshmukh}
\orcid{0000-0003-4683-5540}
\email{jdeshmuk@usc.edu}
\affiliation{%
  \institution{University of Southern California}
  \city{Los Angeles}
  \state{California}
  \country{USA}
}

\author{Lars Lindemann}
\orcid{0000-0003-3430-6625}
\email{llindema@usc.edu}
\affiliation{%
  \institution{University of Southern California}
  \city{Los Angeles}
  \state{California}
  \country{USA}
}


\renewcommand{\shortauthors}{Trovato et al.}


\begin{abstract}

Multi-agent systems (MASs) consisting of a number of autonomous agents that
communicate, coordinate, and jointly sense the environment to achieve complex
missions can be found in a variety of applications such as robotics, smart
cities, and internet-of-things applications. Modeling and monitoring MAS
requirements to guarantee overall mission objectives, safety, and reliability is
an important  problem. Such requirements implicitly require reasoning
about diverse sensing and communication modalities between agents, analysis of
the dependencies between agent tasks, and the spatial or virtual distance
between agents. To capture such rich MAS requirements, we model agent
interactions via multiple directed graphs, and introduce a new logic -- {\em
Spatio-Temporal Logic with Graph Operators} (STL-GO). The key innovation in
STL-GO are graph operators that enable us to reason about the number of agents
along either the incoming or outgoing edges of the underlying interaction graph
that satisfy a given property of interest; for example, the requirement that an
agent should sense at least two neighboring agents whose task graphs indicate
the ability to collaborate. We then propose novel distributed monitoring
conditions for individual agents that use only local information to determine
whether or not an STL-GO specification is satisfied. We compare the expressivity
of STL-GO against existing spatio-temporal logic formalisms, and demonstrate the
utility of STL-GO and our distributed monitors in a bike-sharing and a
multi-drone case study.

\end{abstract}




\maketitle
\thispagestyle{empty}
\pagestyle{plain}

\section{Introduction}

Multi-agent systems (MASs) are self-organizing systems that consist of autonomous agents that interact and coordinate with each other to achieve individual and collaborative  objectives. MASs can be found in a variety of areas and applications, including robotics \cite{arai2002advances, yan2013survey}, traffic and transportation \cite{burmeister1997application,malikopoulos2021optimal}, distributed control \cite{mesbahi2010graph,bullo2009distributed,dimarogonas2011distributed}, and smart cities \cite{roscia2013smart}.  Compared to single-agent systems, MASs face additional design challenges since: (1) individual agents have limited information about the states of other agents, (2) agents may be coupled via their objectives, (3) system requirements are usually complex and difficult to formulate, even for domain experts, and  (4) verification and control algorithms scale with the number of agents. We thus pose the following question that we are aiming to address in this paper. How should we specify
mission and safety objectives of autonomous MAS, especially when agents interact via multiple network topologies, and how can we efficiently check if these mission and safety objectives are satisfied?

Temporal logics provide a universal tool for expressing complex system properties. In particular,  linear temporal logic (LTL) \cite{eisner2003reasoning} and signal temporal logic (STL) \cite{maler2004monitoring, bartocci2018specification} have been used across different domains. Indeed, there exists a plethora of formal verification and control design techniques that enforce a desired system behavior described by a  temporal logic specification \cite{belta2017formal,baier2008principles,lindemann2025formal}. However, temporal logics do not allow  to reason over spatial properties across different agents in the context of MASs. Therefore, recent works have proposed more expressive spatio-temporal logics that can reason over complex MAS behavior \cite{ma2020sastl, haghighi2015spatel, nenzi2015specifying, nenzi2015qualitative, bartocci2017monitoring, nenzi2022logic}. 
Specifically, spatial aggregation signal temporal logic (SaSTL) extends STL by introducing two additional operators that describe spatial aggregation and spatial counting across sets of agents, which is commonly found in smart cities \cite{ma2020sastl}. Signal spatio-temporal logic (SSTL), on the other hand, was designed to include spatial information in the temporal logic \cite{nenzi2015specifying, nenzi2015qualitative}. Lastly, spatio-temporal reach and escape logic (STREL) employs spatial reach and escape operators, enhancing modeling of more complex spatio-temporal requirements \cite{bartocci2017monitoring, nenzi2022logic}. However, these spatio-temporal logics do not allow to quantitatively reason over multiple network topologies, i.e., when asymmetric task dependencies exist and when diverse sensing and communication modalities are used as we demonstrate later.

Agent interactions in MASs are usually modeled by discrete graphs. Indeed, such graphs play a fundamental role in describing how information flows between agents, how agents perceive each other, how agents share resources with each other, or how agents may be coupled via shared objectives. While most of the existing works focus on a single graph, we would like to explicitly reason over multiple graphs to fully capture these complex dependencies and interactions. For instance, we would like to consider one graph each for communication, sensing, task dependencies, and relative distances. To give a concrete example, we may want to specify that all agents that are in close proximity should either be able to communicate or sense each other. This would involve a distance graph, a communication graph, and a sensing graph, see Figure \ref{fig:fire} from Section \ref{sec:example} where three graphs are shown in Figures \ref{fig:fire} (b), (c), and (d), respectively. Consequently, we argue that spatio-temporal logics that incorporate and quantitatively reason over different graphs are of great importance. In this work, we propose Spatio-Temporal Logic with Graph Operators (STL-GO) and distributed monitoring algorithms for MAS that are described by multiple graphs. Our specific contributions are as follows: 
\begin{itemize}[leftmargin=*]
    \item We propose a two level logical hierarchy where the outer logic (STL-GO) reasons over all agents simultaneously while the inner logic (STL-GO-S) reasons over specific agents and their agent interactions. Our main innovation,  the graph operators ``incoming'' and ``outgoing'', can explicitly reason over multiple asymmetric graphs.
    \item We propose novel distributed monitoring conditions and a distributed monitor for  individual agents that use only local information to determine the satisfaction of an STL-GO-S specification.
    \item We compare STL-GO with STL, SaSTL, and SSTL and show that STL-GO can express counting and certain distance properties that can be expressed in these logics.
    \item We illustrate  STL-GO  and the efficiency of our distributed monitoring algorithms in  a bike-sharing  and a multi-drone case study.
\end{itemize}

\subsection{Related Work}
Commonly used temporal logics include Computation Tree Logic (CTL) \cite{emerson1982using},  Linear Temporal Logic (LTL) \cite{eisner2003reasoning}, and Signal Temporal Logic (STL) \cite{maler2004monitoring, bartocci2018specification}. In the context of MAS, researchers have proposed various extensions such as Counting LTL \cite{nilsson2016control}, Consensus STL \cite{xu2016census}, and Capability Temporal Logic \cite{leahy2022fast}.
These extensions provide additional flexibility in expressing  rich MAS behaviors.

Other works have focused on expressing requirements for MAS using  Alternating-Time Temporal Logic (ATL) \cite{goranko2004comparing, dimitrova2014deductive} and its extended version in \cite{beutner2021temporal}. Embedded graph grammars have been used to capture timed and spatial interactions between agents \cite{smith2009multi, mcnew2007solving, guo2016hybrid}.
Graph Temporal Logic (GTL) \cite{xu2019graph} describes tasks that involve inferring spatial-temporal logic formulas from data on labeled graphs.
To address the spatial components of multi-agent systems, spatio-temporal logics have been introduced as discussed before, e.g., SpaTeL \cite{haghighi2015spatel}, STREL \cite{bartocci2017monitoring, nenzi2022logic}, SaSTL \cite{ma2020sastl}, SSTL \cite{nenzi2015specifying, nenzi2015qualitative}. We note that we provide a detailed comparison later in Section \ref{sec:exp}.
To account for uncertainty in these settings, probabilistic extensions like PCTL \cite{hansson1994logic}, PrSTL \cite{sadigh2016safe}, and C2TL \cite{jha2018safe} have been developed. 
To address challenges in managing complex topologies and enhancing runtime scalability, scenario-based hierarchical modeling languages for reconfigurable CPSs have been proposed  \cite{wang2024scenario}.
To address the challenges of multi-rate control systems, Multiclock Logic (MCL) \cite{incer2024layered} enables the formal specification of requirements for individual control components from the perspective of their local clocks.
However, existing frameworks still lack comprehensive integration and reasoning capabilities over multiple asymmetric interaction graphs.

On the monitoring side, recent work has expanded STL monitoring to handle complex specifications and dynamic environments across temporal and spatial domains.
Offline STL monitoring methods analyze behaviors post-execution \cite{maler2004monitoring, bartocci2018specification}, while robust online monitoring handles real-time data with partial information
\cite{deshmukh2017robust} and addresses uncertainties \cite{finkbeiner2022truly}.
Predictive monitoring provides monitoring results for partial information by anticipating future behaviors \cite{qin2019predictive, lindemann2023conformal, zhao2024robust, ma2021predictive}, and specialized techniques like online causation and reset monitoring enhance adaptability \cite{zhang2023online, zhang2022online}. 
For multi-agent systems, existing spatio-temporal logics monitoring frameworks \cite{bartocci2017monitoring, nenzi2022logic, ma2020sastl, nenzi2015specifying, nenzi2015qualitative, zhao2025distributionally} monitor complex interaction behaviors in a centralized way.
Recent work also investigates distributed monitoring approaches that leverage partially synchronous settings and SMT-based techniques for STL monitoring in distributed cyber-physical systems \cite{momtaz2023predicate, momtaz2023monitoring}.

\section{Multi-Agent Interactions over Multiple Network Topologies}

We consider a MAS consisting of $N$ agents, where we denote the set of agents as $\mathcal{V} \coloneqq \{1, \dots, N\}$.
We use $\mathbb{T}$ to denote the time domain, where we particularly consider continuous and discrete time with $\mathbb{T} \coloneqq \mathbb{R}_{\ge 0} \cup \{+\infty\}$  and $\mathbb{T} \coloneqq \mathbb{N} \cup \{+\infty\}$, respectively.
A trajectory of agent $i$ is a function $\mathbf{x}^i: \mathbb{T} \to \mathbb{R}^n$, and we denote $x_t^i \in \mathbb{R}^n$ as the state of agent $i$ at time $t$. 
The MAS trajectory is an $N$-tuple $\mathbf{x} \coloneqq (\mathbf{x}^1, \dots, \mathbf{x}^N)$ and the MAS state at time $t$ is $x_t \coloneqq (x^1_t, \dots, x^N_t)$.
For simplicity, we assume that the MAS is homogeneous, i.e., each agent’s state has the same dimension. 
We model agent interactions by discrete multigraphs, allowing multiple edges to connect the same pair of nodes. A multigraph is defined as
$\mathcal{G} \coloneqq (\mathcal{V}, \mathcal{E}, w)$, where $\mathcal{E} \subseteq \mathcal{V} \times \mathcal{V} \times \mathbb{N}$ is a set of directed (or undirected) relations (edges) that indicate interactions between agents, e.g., $(i, j, u) \in \mathcal{E}$ is the $u$-th edge between agent $i$ and $j$. The function $w: \mathcal{E} \to \mathbb{R} \cup\{\infty, -\infty\}$ assigns a weight to each edge, e.g., a cost or a distance. 
This multigraph can be simplified to a graph with unique edges (referred to as single-edge graph) by restricting $\mathcal{E}$ to  $\mathcal{E} \subseteq \mathcal{V} \times \mathcal{V} \times \{1\}$, or simply defining $\mathcal{E} \subseteq \mathcal{V} \times \mathcal{V}$.

Importantly, a MAS may interact in ways that we need to capture via multiple graphs, describing different network topologies. STL-GO, defined in the next section, will consider multiple graphs and is distinct from existing spatio-temporal logics. We are thus not limited to a specific number of graphs. In the examples of this paper, we  use a communication graph $\mathcal{G}^c \coloneqq (\mathcal{V}, \mathcal{E}^c, w^c)$ (e.g., radio communication), a sensing graph $\mathcal{G}^s \coloneqq (\mathcal{V}, \mathcal{E}^s, w^s)$ (e.g., camera sensors), a mission dependency graph $\mathcal{G}^{m} \coloneqq (\mathcal{V}, \mathcal{E}^{m}, w^{m})$ indicating collaborative missions, a distance graph $\mathcal{G}^d \coloneqq (\mathcal{V}, \mathcal{E}^d, w^d)$ indicating relative agent distances, and a shortest distance graph $\mathcal{G}^{d_s} \coloneqq (\mathcal{V}, \mathcal{E}^{d}, w^{d_s})$. Each graph can be directed or undirected, with edge weights indicating communication quality, sensing reliability, relative distances, mission importance, and weight assigned to the shortest path, respectively. These graphs may change over time, and we denote a specific graph at time $t$ by $\mathcal{G}_t^{\text{type}_i} \coloneqq (\mathcal{V}_t^{\text{type}_i}, \mathcal{E}_t^{\text{type}_i}, w_t^{\text{type}_i})$, where ${\text{type}_i} \in \mathcal{T}$ and $\mathcal{T}$ is the set of all graph types with $M:=|\mathcal{T}|$, e.g., $\mathcal{T} \coloneqq \{c, s, m, d, d_s\}$ and $M=5$ in our examples. The time evolution (or trajectory) of a graph is expressed as a function $\pmb{\mathcal{G}}^{\text{type}_i}: \mathbb{T} \to \mathcal{G}^{\text{type}_i}$.
The trajectory of all graphs is denoted by an $M$-tuple $\pmb{\mathcal{G}} \coloneqq (\pmb{\mathcal{G}}^{\text{type}_1}, \dots, \pmb{\mathcal{G}}^{\text{type}_M})$, where the graph at time $t$ is $\mathcal{G}_t \coloneqq (\mathcal{G}^{\text{type}_1}_t, \dots, \mathcal{G}^{\text{type}_M}_t)$ with $\text{type}_i \in \mathcal{T}$ for $i \in \{1, \dots, M\}$.
In the following, we will use the simplified notation $\mathcal{G}^{\text{type}}$ instead of $\mathcal{G}^{\text{type}_i}$.
Finally, we compactly write the MASs state and graph as $\mathcal{MA} \coloneqq (\mathbf{x}, \pmb{\mathcal{G}})$.


\section{STL-GO: Spatio-Temporal Logic with Graph Operators}\label{sec:stlg}
We now  define the syntax and semantics of Spatio-Temporal Logic with Graph Operators (STL-GO). 
We first present STL-GO-S in Section \ref{subsec:inner} to reason over individual agents and their agent interactions, i.e., 
STL-GO-S focuses on spatio-temporal tasks imposed on an individual agent. For example, agent $i$ should visit a specific location while being able to communicate with at least two neighboring agents. We then present STL-GO in Section \ref{subsec:outer} to reason over STL-GO-S tasks imposed on multiple agents, i.e., STL-GO focuses on spatio-temporal tasks imposed on potentially all agents. For example, there should exist at least one agent that visits a specific location while being able to communicate with at least two other agents.

\subsection{STL-GO-S: STL-GO for Individual Agents}\label{subsec:inner}

STL-GO-S uses predicates $\pi^{\mu_x}:\mathbb{R}^n \to \mathbb{B}$, where  $\mathbb{B} \coloneq \{\top, \bot\}$ is the set of true and false, to express atomic constraints over the state $x_t^i$ of a single agent. The truth value of $\pi^{\mu_x}$ is determined by a predicate function $\mu_x: \mathbb{R}^n \to \mathbb{R}$, i.e., $\pi^{\mu_x}(x_t^i)=\top$ iff $\mu_x(x_t^i)\ge 0$. The syntax of STL-GO-S is
\begin{align}
    \varphi ::= \top \mid \pi^{\mu_x} \mid \neg \varphi \mid \ & \varphi_1 \wedge \varphi_2 \mid \varphi_1 \mathbf{U}_{I} \varphi_2 \mid \mathbf{In}^{W, \#}_{\mathcal{G}, E}  \varphi \mid\mathbf{Out}^{W, \#}_{\mathcal{G}, E} \varphi  \nonumber
\end{align}
where the first five rules are the same as in STL \cite{maler2004monitoring, bartocci2018specification}, while the last two rules define our novel graph operators.
Specifically, the operators $\neg$ and $\wedge$ are the standard Boolean ``negation'' and ``conjunction'', respectively, which can further induce ``disjunction'' by $\varphi_1 \vee \varphi_2:=\neg(\neg \varphi_1 \wedge \neg \varphi_2)$ and ``implication" by $\varphi_1 \to \varphi_2:= \neg \varphi_1 \vee \varphi_2$.
The operator $\mathbf{U}_{I}$ is the temporal operator ``until'', where $I \coloneqq [t_1, t_2]$ is a time interval where $t_1, t_2 \in \mathbb{R}_{\ge 0}$ with $t_1 \le t_2$.
The temporal operators ``eventually'' and ``always'' can be derived from ``until'' by $\mathbf{F}_{I} \varphi:= \top \mathbf{U}_{I} \varphi$ and $\textbf{G}_{I} \varphi:=\neg \mathbf{F}_{I} \neg \varphi$, respectively.

The operators $\mathbf{In}^{W, \#}_{\mathcal{G}, E}$ and $\mathbf{Out}^{W, \#}_{\mathcal{G}, E}$ are the graph operators ``incoming'' and ``outgoing'', respectively, where $E \coloneqq [e_1, e_2]$ with $e_1, e_2 \in \mathbb{N}\cup \{0, +\infty\}$ and $e_1 \leq e_2$ is an interval that constrains the number of (incoming or outgoing) edges that satisfy a property, while $W \coloneqq [w_1, w_2]$ with $w_1, w_2 \in \mathbb{R} \cup \{-\infty, +\infty\}$ and $w_1 \leq w_2$ is an interval that constrains the corresponding weights. An STL-GO-S formula $\varphi$ is hence not only related to the trajectory of an individual agent $i$ but also to the trajectories of neighboring agents, as described by a set of graphs $\mathcal{G}\subseteq \mathcal{T}$. We use the notation $(\mathcal{MA}, i, t) \models \varphi$ to denote the STL-GO-S semantics which define when an STL-GO-S formula $\varphi$ imposed on agent $i$ at time $t$ is satisfied. These semantics follow standard rules for the STL operators, as defined below. However,  the semantics of the graph operators are defined as
\begin{align*}
    (\mathcal{MA}, &i, t) \models \mathbf{In}^{W, \#}_{\mathcal{G}, E} \varphi  \;  \text{iff} \; \# \ \mathcal{G}^\text{type} \in \mathcal{G}\text{ s.t. } |\{(j, i, n) \in \mathcal{E}_t^\text{type}  \mid w_t^\text{type}(j, i, n) \in W  \wedge (\mathcal{MA}, j, t) \models \varphi \}| \in E, \\
    (\mathcal{MA}, &i, t) \models \mathbf{Out}^{W, \#}_{\mathcal{G}, E} \varphi  \;  \text{iff} \; \# \ \mathcal{G}^\text{type} \in \mathcal{G}\text{ s.t. }  |\{(i, j, n) \in \mathcal{E}_t^\text{type}  \mid w_t^\text{type}(i, j, n) \in W  \wedge (\mathcal{MA}, j, t) \models \varphi \}| \in E, 
\end{align*}
where $\# \in \{\exists, \forall\}$ can be either $\exists$ (existential) or $\forall$ (universal).  Consequently, depending on the choice of $\#$ in $\mathbf{In}^{W, \#}_{\mathcal{G}, E}$ and $\mathbf{Out}^{W, \#}_{\mathcal{G}, E}$, we obtain two modes of the graph operator.\footnote{The $\exists$ and $\forall$ quantifiers over
graphs could hypothetically be replaced with disjunction and conjunction operators
over  graphs. However, this adds to the representational complexity of the formula. More economic encodings are generally favorable, e.g., \cite{french2013succinctness} argues in favor of the $\exists$ and $\forall$ quantifiers   via different notions of representational succinctness for multi-modal logics over graphs.} We also remark that   we interpret the cardinality of the empty set as $|\emptyset|=0$.

The existential incoming operator $\mathbf{In}^{W, \exists}_{\mathcal{G}, E} \varphi$ expresses that there exists one graph (i.e., $\exists\mathcal{G}^\text{type}\in \mathcal{G}$) for which the number of edges $(j, i, n)$ 
that provide information from agent $j$ to agent $i$ via $\mathcal{E}_t^\text{type}$ (i.e., $(j, i, n) \in \mathcal{E}_t^\text{type}$) are within the interval $E$ and satisfy the following two conditions: (1) the weight of the $n$th edge from agent $j$ to agent $i$ is within $W$ (i.e., $w_t^\text{type}(j, i, n) \in W$), and (2) the agent $j$ should satisfy $\varphi$ (i.e., $(\mathcal{MA}, j, t) \models \varphi$). Similarly, the universal incoming operator $\mathbf{In}^{W, \forall}_{\mathcal{G}, E} \varphi$ requires that the same property holds, but now for all graphs (i.e., $\forall\mathcal{G}^\text{type}\in\mathcal{G}$).
On the other hand, the existential outgoing operator $\mathbf{Out}^{W,\exists}_{\mathcal{G}, E} \varphi$ expresses that there exists one graph (i.e., $\mathcal{G}^\text{type}\in\mathcal{G}$) for which the number of edges $(i, j, n)$ that provide information from agent $i$ to agent $j$ via $\mathcal{E}_t^\text{type}$ (i.e., $(i, j, n) \in \mathcal{E}_t^\text{type}$) are within the interval $E$ and satisfy the following two conditions: (1) the weight of the $n$th edge from agent $i$ to $j$ is within $W$ (i.e., $w_t^\text{type}(i, j, n) \in W$), and (2) the agent $j$ should satisfy $\varphi$ (i.e., $(\mathcal{MA}, j, t) \models \varphi$). The interpretation of the universal outgoing operator follows analogously. If we are not interested in weights, i.e., $W:= [-\infty, \infty]$, we simply write  $\mathbf{In}^{\#}_{\mathcal{G}, E} \varphi$ and  $\mathbf{Out}^{\#}_{\mathcal{G}, E} \varphi$.

We note that the difference between the incoming and outgoing operators are the direction of the information flow to and from agent $i$, respectively. We further remark that the incoming and outgoing operators are the same if the graphs in $\mathcal{G}$ are undirected. Note also that the concatenation of these graph operators can be used to describe information flow over multiple hops. In the special case where we only consider a single-edge graph (i.e., $|\mathcal{G}|=1$ and $(i,j,u)\in\mathcal{E}$ with $u=1$), the semantics of these two operators reduce to $(\mathcal{MA}, i, t) \models \mathbf{In}^{W}_{\mathcal{G}, E} \varphi  \  \text{iff} \ |\{j \in \mathcal{V} \mid w_t(j, i, 1) \in W \wedge (\mathcal{MA}, j, t) \models \varphi \}| \in E$ and $(\mathcal{MA}, i, t) \models \mathbf{Out}^{W}_{\mathcal{G}, E} \varphi  \  \text{iff} \ |\{j \in \mathcal{V} \mid w_t(i, j, 1) \in W \wedge (\mathcal{MA}, j, t) \models \varphi \}| \in E$. If additionally the graph is undirected,  we obtain a definition similar to the counting operator in~\cite{ma2020sastl}. 


Lastly, we define the semantics of the remaining Boolean and temporal operators, which follow standard convention \cite{maler2004monitoring, bartocci2018specification} as
\begin{align}\label{eq:inner_semantics}
    (\mathcal{MA}, i, t) \models \top  \ & \text{iff} \ \text{True},  \nonumber  \\
    (\mathcal{MA}, i, t) \models \pi^{\mu_x}  \ & \text{iff} \ \mu(x_t^i) \ge 0, \nonumber \\
    (\mathcal{MA}, i, t) \models \neg \varphi  \ & \text{iff} \ (\mathcal{MA}, i, t) \not\models \varphi, \nonumber \\
    (\mathcal{MA}, i, t) \models \varphi_1 \wedge \varphi_2  \ & \text{iff} \ (\mathcal{MA}, i, t) \models \varphi_1 \wedge (\mathcal{MA}, i, t) \models \varphi_2, \nonumber \\
    (\mathcal{MA}, i, t) \models \varphi_1 \mathbf{U}_{I} \varphi_2  \ & \text{iff} \ \exists t' \in  (t \oplus I)\cap \mathbb{T},  (\mathbf{x}^i, t') \models \varphi_2 \ \text{and} \nonumber  \forall t'' \in [t, t']\cap \mathbb{T}, (\mathbf{x}^i, t'') \models \varphi_1, \nonumber 
\end{align}
where $t \oplus I \coloneqq \{t + t_1 \mid t_1 \in I \}$ is the Minkowski sum.

The primary advantage of STL-GO-S is its ability to describe properties defined over multiple asymmetric multigraphs, a feature not supported by existing  spatio-temporal logics \cite{ma2020sastl, nenzi2015specifying, nenzi2015qualitative, bartocci2017monitoring, nenzi2022logic}. We illustrate the expressive strength of STL-GO-S in two examples.

\begin{myexm}[Explicit redundancy requirements over different graphs]\label{exm:redundancy1}
We would like to express that, at all times, there is at least one agent that can sense or communicate with agent $i$. Therefore, we consider the formula  $\varphi \coloneqq \mathbf{G}_{[0, +\infty]} (\mathbf{In}^{\exists}_{\{\mathcal{G}^s, \mathcal{G}^c\}, [1, +\infty]} \top)$
    where the sensing and communication graphs are undirected and  single-edged. In this way, we  express a redundancy requirement for cases where sensing or communication links may fail.
    For directed graphs, and if we want that at least one agent can sense or communicate with agent $i$, or vice versa, we instead write $\varphi:=\mathbf{G}_{[0, +\infty]} (\mathbf{In}^{\exists}_{\{\mathcal{G}^s,\mathcal{G}^c \}, [1, +\infty]} \top \vee \mathbf{Out}^{\exists}_{\{\mathcal{G}^s, \mathcal{G}^c\}, [1, +\infty]} \top)$, see Section \ref{subsec:drone} for similar examples.
\end{myexm}

\begin{myexm}[Multigraphs in web applications]
    Let the MAS be a system of web servers hosting an arbitrary number of clients. Consider a communication graph $\mathcal{G}^c$  where each incoming edge to a server is an HTTP request from a client. An outgoing edge from a server denotes an HTTP response to a client. Multiple requests can be served in a single connection so that $\mathcal{G}^c$ is a multigraph. We are interested in protecting the server from overloading, and thus consider the formula $\varphi \coloneqq \mathbf{G}_{[0, +\infty]} (\mathbf{In}^{\exists}_{\{\mathcal{G}^c\}, [0, \zeta]}\top)$ where $\zeta \in \mathbb{N}$ is a request limit. We could further specify the type of clients sending the requests, in which case we replace $\top$ in $\varphi$ with $x = b$ where $x$ is the state of a client and $b \in \mathbb{N}$ denotes the type.
\end{myexm}


\subsection{STL-GO for Multiple Agents}\label{subsec:outer}

We now introduce STL-GO to reason over multiple agents simultaneously by building up on  STL-GO-S. The syntax of STL-GO is defined as
\begin{align}
    \phi ::= \top \mid \pi^{\boldsymbol{\mu}_x} \mid i.\varphi \mid \neg \phi \mid \phi_1 \wedge \phi_2 \mid \phi_1 \mathbf{U}_{I} \phi_2, \nonumber 
\end{align}
where $\varphi$ denotes an STL-GO-S formula and $\phi$ denotes an STL-GO formula. The operators $\top$, $\neg$, $\wedge$, and $\mathbf{U}_{I}$ are the same as before. However, STL-GO uses predicates $\pi^{\boldsymbol{\mu}_x}:\mathbb{R}^{n \cdot N}\to \mathbb{B}$, to express atomic constraints over the MAS state $x_t$. The truth value of $\pi^{\boldsymbol{\mu}_x}$ is determined by the sign of the predicate function $\boldsymbol{\mu}_x: \mathbb{R}^{n \cdot N} \to \mathbb{R}$, i.e., $\pi^{\boldsymbol{\mu}_x}(x_t)=\top$ iff $\boldsymbol{\mu}_x(x_t)\ge 0$.
We distinguish predicate functions in STL-GO-S and STL-GO by $\mu_x$ and $\boldsymbol{\mu}_x$, respectively, and we note  that $\pi^{\mu_x}$ in STL-GO-S describes a predicate over one agent (the input is $x_t^i$), while $\pi^{\boldsymbol{\mu}_x}$ in STL-GO describes a predicate over all agents (the input is $x_t$). Lastly, we introduce a new operator $i.\varphi$ to couple an STL-GO-S formula $\varphi$ imposed on agent $i$ into STL-GO.

We use the notation $(\mathcal{MA}, t) \models \phi$ to denote that the STL-GO formula $\phi$ is satisfied by the MAS at time $t$. Formally, the semantics of $\phi$ are  inductively defined as 
\begin{align}
    (\mathcal{MA}, t) \models \top  \ & \text{iff} \ \text{True},  \nonumber  \\
    (\mathcal{MA}, t) \models \pi^{\boldsymbol{\mu}_x}  \ & \text{iff} \ \boldsymbol{\mu}({x}_t) \ge 0, \nonumber \\
    (\mathcal{MA}, t) \models i.\varphi  \ & \text{iff} \ (\mathcal{MA}, i, t) \models \varphi, \nonumber \\
    (\mathcal{MA}, t) \models \neg \phi  \ & \text{iff} \ (\mathcal{MA}, t) \not\models \phi, \nonumber \\
    (\mathcal{MA}, t) \models \phi_1 \wedge \phi_2  \ & \text{iff} \ (\mathcal{MA}, t) \models \phi_1 \wedge (\mathcal{MA}, t) \models \phi_2, \nonumber \\
    (\mathcal{MA}, t) \models \phi_1 \mathbf{U}_{I} \phi_2  \ & \text{iff} \ \exists t' \in (t \oplus I) \cap \mathbb{T},  (\mathcal{MA}, t') \models \phi_2 \ \text{and}   \forall t'' \in [t, t']\cap \mathbb{T}, (\mathcal{MA}, t'') \models \phi_1. \nonumber 
\end{align}
We  highlight the semantics of the $i.\varphi$ operator, which follows the STL-GO-S semantics for agent $i$. For convenience, we additionally derive a universal and an existential   operator over $i.\varphi$, i.e.,  we define $\textbf{FA}_{V} \varphi := \wedge_{i \in V} i.\varphi$ and  $\textbf{EX}_{V} \varphi := \vee_{i \in V} i.\varphi$ which are equivalent to $\forall i \in V, (\mathcal{MA}, t) \models i.\varphi$ and $\exists i \in V$, such that $(\mathcal{MA}, t) \models i.\varphi$, respectively. STL-GO extends STL-GO-S in two ways: (1) it allows for the combination of multiple STL-GO-S formulae imposed on different agents, and (2) the predicate $\pi^{\boldsymbol{\mu}_x}$ enables us to reason over atomic constraints involving multiple  agents.

\begin{myexm}[Redundancy for all agents]
\label{exm:redundancy2}
    Example \ref{exm:redundancy1} expressed redundancy requirements in communication and sensing. We would now like to consider a similar requirement, but  for all agents that are close to agent $i$, e.g., to ensure collision avoidance and safety. We thus additionally consider a distance graph.
    Specifically, we would like to express that, at all times,  all agents within a distance of 1 meter from agent $i$ can either sense or communicate with agent $i$. The STL-GO formula here is
    \begin{align}
        \phi \coloneqq \mathbf{G}_{[0, +\infty]} \bigwedge_{j \in \mathcal{V} \setminus \{i\}}  \mathbf{In}^{[0, 1], \exists}_{\{\mathcal{G}^{d,i, j}\}, [1, 1]} \top \implies j.(\mathbf{In}^{\exists}_{\{\mathcal{G}^{s,i},\mathcal{G}^{c,i}\}, [1, +\infty]} \top), \nonumber
    \end{align}   
    where $\mathcal{G}^{d,i, j}$ is a subgraph of $\mathcal{G}^{d}$ only containing agents $i$ and $j$. Similarly, the graphs $\mathcal{G}^{s,i}$ and $\mathcal{G}^{c,i}$ are subgraphs of $\mathcal{G}^{s}$ and $\mathcal{G}^{c}$ only containing agents $i$ and its neighbors. 
    If the state of the agent contains information about the agent's position, we can equivalently write $\phi$ as $\phi \coloneqq \mathbf{G}_{[0, +\infty]} \bigwedge_{j \in \mathcal{V} \setminus \{i\}}  \pi^{\boldsymbol{\mu}_{d,j}} \implies j.(\mathbf{In}^{\exists}_{\{\mathcal{G}^{s,i},\mathcal{G}^{c,i}\}, [1, +\infty]} \top)$, where $\boldsymbol{\mu}_{d,j} = 1 - ||x^i - x^j||_2$ is the predicate function in STL-GO. 
    
\end{myexm}

\begin{myexm}[Leader follower requirements]
    Agents $i$ and $j$ (here considered followers) are supposed to gather information from an information center and relay the information to agent $k$ (here considered the leader) during the time interval $[5, 15]$. Agents $i$ and $j$ thus have to remain within 2 meters distance of agent $k$, which limits its ability to stay close to the information center.
    Therefore, we require that at least one of agent $i$'s or  $j$'s neighbors in  the undirected communication graph is within 1 meter distance of the information center, with the additional requirement that the  communication weight should be larger than 10. The STL-GO formula is 
    \begin{align}
        \phi \coloneqq \mathbf{G}_{[5, 15]} \Big( \pi^{\boldsymbol{\mu}_x} \wedge \textbf{EX}_{\{i, j\}} \big( \mathbf{In}^{[10, +\infty], \exists}_{\{\mathcal{G}^c\}, [1, +\infty]} \pi^{\mu_x}  \big) \Big), \nonumber
    \end{align}
    with $\boldsymbol{\mu}_x(\mathbf{x}) = \min(2 - ||x^i - x^k||_2, 2 - ||x^j - x^k||_2)$ and $\mu_x(x) = 1 - ||x - x^{center}||_2$, where $x^{center}$ is the position of the information center.
\end{myexm}

\textbf{Formula length. } The satisfaction of an STL-GO-S formula $\varphi$ depends on the states and graphs within a specific time interval, known as its horizon.
This horizon of $\varphi$ is denoted by $[S_{\varphi}, T_{\varphi}]$, where $S_{\varphi}$ and $T_{\varphi}$ are the minimum and maximum time instants needed to decide if $\varphi$ is satisfied. They are computed recursively on the structure of $\varphi$ and follows standard computation for Boolean and temporal operators, see e.g., \cite{dokhanchi2014line}, while graph operators  do not affect $S_{\varphi}$ and $T_{\varphi}$. Formally, we have
\begin{align*}
    S_{\top} &= T_{\top} = 0\\
    S_{\pi^{\mu_x}} &= T_{\pi^{\mu_x}}  = 0\\
    S_{\neg \varphi_1} &= S_{\varphi_1}, T_{\neg \varphi_1} = T_{\varphi_1}\\
    S_{\varphi_1 \wedge \varphi_2} &= \min(S_{\varphi_1},  S_{\varphi_2}), T_{\varphi_1 \wedge \varphi_2} = \max(T_{\varphi_1}, T_{\varphi_2})\\
    S_{\varphi_1 \mathbf{U}_{[a,b]} \varphi_2} &= a + \min(S_{\varphi_1}, S_{\varphi_2}), T_{\varphi_1 \mathbf{U}_{[a,b]} \varphi_2} = b + \max(T_{\varphi_1}, T_{\varphi_2})\\
    S_{\mathbf{In}^{W, \#}_{\mathcal{G}, E} \varphi_1} &= S_{\varphi_1}, T_{\mathbf{In}^{W, \#}_{\mathcal{G}, E} \varphi_1} = T_{\varphi_1}\\
    S_{\mathbf{Out}^{W, \#}_{\mathcal{G}, E} \varphi_1} &= S_{\varphi_1}, T_{\mathbf{Out}^{W, \#}_{\mathcal{G}, E} \varphi_1} = T_{\varphi_1}.
\end{align*}

The horizon of an STL-GO formula $\phi$ is defined as $[S_{\phi}, T_{\phi}]$ and calculated the same way as for STL-GO-S with the addition of  $S_{i.\varphi} = S_{\varphi}$ and $T_{i.\varphi} = T_{\varphi}$ for the operator $i.\varphi$.

\subsection{Centralized Offline Monitoring under STL-GO}\label{subsec:centralized}

While STL-GO is defined over a general time domain $\mathbb{T}$, we focus on the discrete-time setting in the remainder of the paper, i.e., $\mathbb{T} = \mathbb{N}$.
We will first look at a centralized offline monitoring problem, (i.e., given global knowledge of the MAS trajectories of $\mathbf{x}$ and $\pmb{\mathcal{G}}$, we want to check whether or not an STL-GO formula $\phi$ is satisfied), while we look at the distributed setting in the next section.



\begin{myprob}
    Given an STL-GO formula $\phi$, discrete-time trajectories of agents and graphs $\mathcal{MA} \coloneqq (\mathbf{x}, \pmb{\mathcal{G}})$, and monitoring time $T$, determine a Boolean satisfaction signal $s_{\phi}: [0, T] \to \{0,1\}$, such that $s_{\phi}(t) = 1$ if $(\mathcal{MA}, t) \models \phi$ and $s_{\phi}(t) = 0$ otherwise. 
\end{myprob}

Given a formula $\phi$, we compute $s_\phi$ recursively by  applying the semantics to the structure of $\phi$. Note that the structure of a formula $\phi$ can be thought of as a tree, e.g., the formula $\varphi \coloneqq \mathbf{G}_{[0, +\infty]} (\mathbf{In}^{\exists}_{\{\mathcal{G}^s, \mathcal{G}^c\}, [1, +\infty]} \top)$ from Example \ref{exm:redundancy1} has  the operator $\mathbf{G}_{[0, +\infty]}$ as a root node  and $\top$ as a leaf node. Specifically, for every subformula $\phi'$ in an STL-GO formula $\phi$, we construct a Boolean signal $s_{\phi'}: [0, T + T_\phi] \to \{0,1\}$, while for every subformula $\varphi'$ in an STL-GO-S formula $\varphi$  and for every agent $i \in \{1, \dots, N\}$ (or a subset of all agents), we construct a Boolean signal $s_{\varphi', i}: [0, T+T_\phi] \to \{0,1\}$.  We then compose the monitors for each subformula bottom-up from the leaf node. 

\textbf{STL-GO-S. }The computation of the monitoring signal $s_{\varphi', i}$  for \textsf{True}, predicates, negation, conjunction, and until follows directly from the definition of the semantics and has previously been studied, see e.g., \cite{maler2004monitoring, bartocci2018specification}. The computation of the monitoring signal $s_{\varphi', i}$ associated with graph operators is more interesting. We consider the case where $\mathcal{G}$ is a single graph, i.e., $|\mathcal{G}|=1$, in which case the existential and universal graph operators are equivalent. The extension to multiple graphs is straightforward and omitted for brevity.\footnote{We simply treat $\exists\mathcal{G}^\text{type} \in \mathcal{G}$ in the definitions of $\mathbf{In}^{W, \exists}_{\mathcal{G}, E}$ and $\mathbf{Out}^{W, \exists}_{\mathcal{G}, E}$  as $|\mathcal{G}|-1$ disjunctions and $\forall\mathcal{G}^\text{type} \in \mathcal{G}$ in the definitions of $\mathbf{In}^{W, \forall}_{\mathcal{G}, E}$ and $\mathbf{Out}^{W, \forall}_{\mathcal{G}, E}$ as $|\mathcal{G}|-1$ conjunctions. } We denote agent $i$'s  set of neighbors with direction $d \in \{\mathbf{In}, \mathbf{Out}\}$ within the weight interval $W$ at time $t$ by $\mathcal{N}^{W, d}_{\mathcal{G}, t} (i)$, which is formally defined as
\begin{align}
    \mathcal{N}^{W, d}_{\mathcal{G}, t}(i) \coloneqq 
    \begin{aligned}\
        \begin{cases}
            \{(j, i, n) \in \mathcal{E}_t \mid  w_t(j, i, n) \in W \} \ & \text{if} \ d = \mathbf{In} \\
            \{(i, j, n) \in \mathcal{E}_t \mid w_t(i, j, n) \in W\} \ & \text{if} \ d = \mathbf{Out}
        \end{cases}
    \end{aligned}. 
\end{align}
Then, for the incoming operator $\varphi' \coloneqq \mathbf{In}^W_{\mathcal{G}, E} \varphi_1$, we have
\begin{align}
    s_{\varphi', i}(t) =  \mathbf{1}\Big(\sum_{(j, i, n) \in \mathcal{N}^{W, \mathbf{In}}_{\mathcal{G}, t}(i)} s_{\varphi_1, j}(t) \in E\Big), \nonumber
\end{align}
where $\mathbf{1}(\cdot)$ is the indicator function, e.g., $\mathbf{1}(\sum_{(j, i, n) \in \mathcal{N}^{W, \mathbf{In}}_{\mathcal{G}, t}(i)} s_{\varphi_1, j}(t) \in E) = 1$ if $\sum_{(j, i, n) \in \mathcal{N}^{W, \mathbf{In}}_{\mathcal{G}, t}(i)} s_{\varphi_1, j}(t) \in E$ and $0$ otherwise. 
For the outgoing operator $\varphi' \coloneqq \mathbf{Out}^W_{\mathcal{G}, E} \varphi_1$, we have
\begin{align}
    s_{\varphi', i}(t) =  \mathbf{1}\Big(\sum_{(i, j, n) \in \mathcal{N}^{W, \mathbf{Out}}_{\mathcal{G}, t}(i)} s_{\varphi_1, j}(t) \in E\Big). \nonumber
\end{align}

\textbf{STL-GO. } The computation of the monitoring signal $s_{\phi'}$, which does not contain graph operators, follows directly from the definition of the semantics, see again \cite{maler2004monitoring, bartocci2018specification} for details. However, it is worth mentioning that the computation of the monitoring signal $s_{\phi'}$ for $\phi' = i.\varphi$ is based on the computation of the Boolean signal $s_{\varphi, i}$ for the STL-GO-S formula $\varphi$, i.e., $s_{\phi'}(t) = s_{\varphi, i}(t)$.

\section{Distributed Offline Monitoring under STL-GO-S}\label{sec:distributed}

Global knowledge of $\mathbf{x}$, as assumed in the previous section, may not always be available. We will thus present distributed monitoring algorithms where an agent only uses partial knowledge of $\mathbf{x}$. To be specific, agent $i$ is equipped with its own monitor that uses only information available to agent $i$, i.e., its own state and potentially available information about other agents. We formally denote the state trajectory that is available to agent $i$  as $i.\bar{\mathbf{x}} \coloneqq (i.\bar{\mathbf{x}}^1, \dots, i.\bar{\mathbf{x}}^N)$, where 
$i.\bar{\mathbf{x}}^i = \mathbf{x}^i$ is its own state and $i.\bar{x}^j_t = x^j_t$ is  agent $j$'s state (for $j \neq i$). If the state $x^j_t$ is not known to agent $i$, then we set  $i.\bar{x}^j_t = \bot$ and will not be able to use this information. 
\begin{remark}
    We remark that we keep the definition of $\bar{\mathbf{x}}$ general on purpose. What we mean is that agent $j$'s state  $x_t^j$ could be obtained by agent $i$ through various means. For example, the state may be obtained via the sensing or the communication graphs directly. Additionally, the state could be relayed to agent $i$ by other agents that can observe agent $j$’s state.   
\end{remark}

While agent $i$ uses partial knowledge of $\mathbf{x}$, it requires global knowledge of  $\pmb{\mathcal{G}}$.

\begin{assumption}\label{ass1}
    We assume that agents have knowledge of $\pmb{\mathcal{G}}$.
\end{assumption}

Assumption \ref{ass1} implies that agent $i$ knows the topology of $\pmb{\mathcal{G}}$. However, agent $i$ does not know information that may flow through the network $\pmb{\mathcal{G}}$, as captured by $i.\bar{\mathbf{x}}$. Having knowledge of $\pmb{\mathcal{G}}$ is often a reasonable assumption, e.g., in situations where the network topology is static or where it changes slowly%
\footnote{Note that a static or a slowly-changing network topology is a sufficient but not a necessary condition for Assumption \ref{ass1}.}. 
In the latter case,  a shared understanding of $\pmb{\mathcal{G}}$ can be obtained via communication.

As agent $i$ only uses partial information, a monitor for agent $i$ may not always be able to determine whether or not an  STL-GO-S formula is satisfied. In such cases, we will use the symbol $?$ to represent an undetermined monitoring outcome.

\begin{myprob}\label{prob:dist}
    Given an  STL-GO-S formula $\varphi$,  partial knowledge $i.\bar{\mathbf{x}}$ of ${\mathbf{x}}$,  knowledge of $\pmb{\mathcal{G}}$, and monitoring time $T$, determine a trinary signal $i.s_{\varphi, i}: [0, T+T_\varphi] \to \{0,1\} \cup \{?\}$ such that 
    $(\mathcal{MA}, i, t) \models \varphi$ if $i.s_{\varphi, i}(t) = 1$, $(\mathcal{MA}, i, t) \not\models \varphi$ if $i.s_{\varphi, i}(t) = 0$, and $i.s_{\varphi, i}(t) = ?$ if the monitoring result is undetermined.
\end{myprob}

We remark that the first $i$ in $i.s_{\varphi, i}$ indicates that it is the $i$-th agent monitor based on information $i.\bar{\mathbf{x}}$, while the second $i$ indicates that the monitor is designed for the STL-GO-S formula $\varphi$ imposed on agent $i$.  Our focus is on monitoring STL-GO-S formulae. We first discuss sufficient conditions that result in a true or false  monitoring result in Section \ref{subsec:sufficient}, and we then present the distributed monitor that solves Problem \ref{prob:dist} in Section \ref{subsec:dist_algo}.

\subsection{Sufficient Conditions for Distributed Monitoring}\label{subsec:sufficient}

Let us start from a motivating  example. 
Consider the STL-GO-S formula $\varphi \coloneqq \mathbf{In}_{\{\mathcal{G}_c\}, [2,+\infty]} \pi^{\mu_x}$, which requires that agent $i$ has at least 2 neighbors in the communication graph satisfying the atomic predicate $\pi^{\mu_x}$.
Agent $i$ is able to obtain a determined answer (as opposed to the case $i.s_{\varphi, i}(t) = ?$) for the monitoring problem if it has access to the states of all its neighbors in the communication graph $\mathcal{G}_c$, or it has at most one neighbor in $\mathcal{G}_c$. 
In the first case, agent $i$ can directly check whether the formula is satisfied or violated. In the second case, agent $i$ can claim that the formula is violated, since it does not have enough neighbors to satisfy the formula.

In more complex scenarios where the formula contains nested graph operators, we must consider neighboring agents and their subsequent neighbors along with their state information to obtain sufficient information for monitoring $\varphi$. 
To formalize this concept, we introduce a graph operator tree. This tree represents the relations of graph operators in the formula $\varphi$, and it specifies the necessary number of neighboring layers for each operator. 

 \textbf{Simplifications.} For simplicity, and as in the previous section, we again consider the case where $\mathcal{G}$ is a single graph, i.e., $|\mathcal{G}|=1$. This means that we can suppress the existential and universal quantifiers $\exists$ and $\forall$ from the graph operators, and instead simply write $\mathbf{In}^{W}_{\mathcal{G}, E}$ and $\mathbf{Out}^{W}_{\mathcal{G}, E} $. Without loss of generality, we further assume that negations do not appear immediately before graph operators, since any such negation can be eliminated by changing the counting interval $E$ to its complement $\mathbb{N} \cup \{0, +\infty\} \setminus E$, e.g., $\neg\mathbf{In}^{W}_{\mathcal{G}, E} \varphi = \mathbf{In}^{W}_{\mathcal{G}, \mathbb{N}\cup \{0, +\infty\} \setminus E} \varphi$.

\textbf{Graph operator tree.} We now assign each graph operator $\mathbf{In}^{W}_{\mathcal{G}, E}$ and $\mathbf{Out}^{W}_{\mathcal{G}, E} $ a unique index $p \in \{1, \dots, \alpha\}$, where $\alpha$ is the total number of graph operators in the formula $\varphi$. 
We denote the subformula related to the $p$th graph operator as $\varphi_p$, i.e., $\varphi_p \coloneqq \mathbf{In}^{W_p}_{\mathcal{G}_p, E_p} \varphi'_p $ or $\varphi_p \coloneqq \mathbf{Out}^{W_p}_{\mathcal{G}_p, E_p} \varphi'_p $, where $\mathcal{G}_p$, $E_p$, and $W_p$ are the graph, counting interval, and distance interval related to the $p$th graph operator, respectively.
We denote the direction of the $p$th graph operator as $d_p \in \{\mathbf{In}, \mathbf{Out}\}$. Before formally defining the graph operator tree, we first provide an illustrative example.

\begin{figure}[t]
	\centering
	\begin{tikzpicture}[
        grow=down,  
        level distance=1.0cm,  
        sibling distance=2.5cm,  
        edge from parent/.style={draw, -latex},  
        every node/.style={draw, rectangle, rounded corners, text centered, minimum width=0.9cm, minimum height=0.6cm}  
        ]
      
        \node {$\varphi$}
          child {node {\tiny{$\mathbf{In}^{W_1}_{\mathcal{G}_1, E_1}$}}
          child {node {\tiny{$\pi^{\mu_1}$}}}
          }
          child {node {\tiny{$\mathbf{Out}^{W_2}_{\mathcal{G}_2, E_2}$}}
            child {node {\tiny{$\pi^{\mu_2}$}}}
            child {node {\tiny{$\mathbf{In}^{W_3}_{\mathcal{G}_3, E_3}$}}
            child {node {\tiny{$\top$}}}
            }
          };

        \node[draw=none] at (-4, 0) {Level 0};
        \node[draw=none] at (-4, -1) {Level 1};
        \node[draw=none] at (-4, -2) {Level 2};
        \node[draw=none] at (-4, -3) {Level 3};
    \end{tikzpicture}

	\caption{An example of the graph operator tree.}

	\label{fig:tree}
\end{figure}
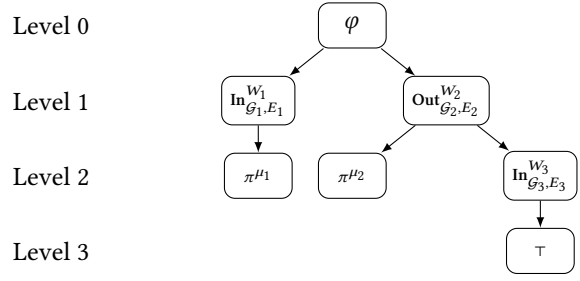

\begin{myexm}\label{exm:tree}
    The graph operator tree of the STL-GO-S formula $\varphi \coloneqq \mathbf{In}^{W_1}_{\mathcal{G}_1, E_1} \pi^{\mu_1}  \mathbf{U}_{I} \mathbf{Out}^{W_2}_{\mathcal{G}_2, E_2} (\pi^{\mu_2} \wedge \mathbf{In}^{W_3}_{\mathcal{G}_3, E_3} \top)$ is shown in Fig.~\ref{fig:tree}.
\end{myexm}

\begin{mydef}
    A graph operator tree for an STL-GO-S formula $\varphi$ is a tree constructed such that: 
    \begin{itemize}[leftmargin=*]
        \item The root node is $\varphi$;
        \item Each intermediate node corresponds to a graph operator. 
        We refer to the intermediate node with  index $p$ as the $p$-th node. 
        For the $p$th node, 
        its child nodes include the standard STL formula $\varphi_{p_s}$ which contains no graph operators (and are hence hidden in our graphical depiction in Figure \ref{fig:tree}), and the graph operators with indices $p_c \in \{1, \dots, \alpha\} \setminus p$, such that $\varphi_p'$ can be reconstructed by $\varphi_{p_s}$ and subformulas $\varphi_{p_c}$ combined through Boolean and temporal operators. 
        \item Each leaf node corresponds to a standard STL formula, where a standard STL formula refers to one that does not contain any graph operators.
    \end{itemize}
\end{mydef}

Note that we do not include Boolean and temporal operators in the graph operator tree since our focus is on the information concerning spatial neighbors. However, the entire formula can still be reconstructed from the tree by appropriately combining the Boolean and temporal operators. 

We  assign each leaf node a unique index $q \in \{1, \dots, \beta\}$, where $\beta$ is the total number of leaf nodes in the graph operator tree. We also denote the subformula related to the $q$th leaf node as $\varphi_q^l$, where superscript $l$ stands for leaf. We further define the level of each node as its distance from the root node, with the root node having a level of 0.
For the $p$th subformula, the level is denoted by $l_p$. For the $q$th leaf node, we define the ancestor node list as $a_q \coloneqq (p_1, \dots, p_r)$ with $r = l_q-1$, which is an ordered sequence of ancestor node indices that connect the $q$th leaf node to the root node. Here, $p_s$ with $s \in \{1, \dots, r\}$ represents the index of the ancestor node with level $s$.

\textbf{Sufficient monitoring conditions.} The idea of our monitor is to propagate state information (about agents other than agent $i$) required to evaluate $\varphi$ from the root node to each leaf node. The evaluation of $\varphi_q^l$ within $\varphi$ relies hence on states of agent $i$'s neighbors according to the ancestor node list $a_q$. 
We define the $r$th level neighbors of agent $i$ at time $t$ as
\begin{align}
    \hat{\mathcal{N}}_{q,t}(i) \coloneqq \hat{\mathcal{N}}^{W_{p_r}, d_{p_r}}_{\mathcal{G}_{p_r}, t}(\hat{\mathcal{N}}^{W_{p_{r-1}}, d_{p_{r-1}}}_{\mathcal{G}_{p_{r-1}}, t}(\dots \hat{\mathcal{N}}^{W_{p_1}, d_{p_1}}_{\mathcal{G}_{p_1}, t}(i)\dots)), \nonumber
\end{align}
where $\hat{\mathcal{N}}^{W, d}_{\mathcal{G}, t}(i)$ is defined as
\begin{align}
    \hat{\mathcal{N}}^{W, d}_{\mathcal{G}, t}(i) \coloneqq 
    \begin{aligned}\
        \begin{cases}
            \{ j \mid (j, i, n) \in \mathcal{E}_t \ \text{and}  \ w_t(j, i, n) \in W \} \ & \text{if} \ d = \mathbf{In} \\
            \{ j \mid (i, j, n) \in \mathcal{E}_t \ \text{and} \ w_t(i, j, n) \in W\} \ & \text{if} \ d = \mathbf{Out}
        \end{cases}
    \end{aligned}. \nonumber
\end{align}
Note that $\mathcal{N}^{W, d}_{\mathcal{G}, t}(i)$ defined in Equation (1) is the set of edges and here $\hat{\mathcal{N}}^{W, d}_{\mathcal{G}, t}(i)$ is the set of agents.
The neighbor set of a group of agents  $V \subseteq \mathcal{V}$ is defined as the union of the neighbor sets of each individual agent in $V$, i.e., $\hat{\mathcal{N}}^{W, d}_{\mathcal{G}, t}(V) \coloneqq \bigcup_{i \in V} \hat{\mathcal{N}}^{W, d}_{\mathcal{G}, t}(i)$.


\begin{myexm}
    For the formula in Example~\ref{exm:tree}, let us assign the indices of graph operators as $\varphi_1 \coloneqq \mathbf{In}^{W_1}_{\mathcal{G}_1, E_1} \pi^{\mu_1}$, $\varphi_2 \coloneqq \mathbf{Out}^{W_2}_{\mathcal{G}_2, E_2} (\pi^{\mu_2} \wedge \mathbf{In}^{W_3}_{\mathcal{G}_3, E_3} \top)$, and $\varphi_3 \coloneqq \mathbf{In}^{W_3}_{\mathcal{G}_3, E_3} \top$, and assign the indices of leaf nodes as $\varphi_1^l \coloneqq \pi^{\mu_1}$, $\varphi_2^l \coloneqq \pi^{\mu_2}$, and $\varphi_3^l \coloneqq \top$.
    Then, we have $a_1 = (1)$ with $r = 1$, $a_2 = (2)$ with $r = 1$, and $a_3 = (2, 3)$ with $r = 2$.
\end{myexm}

We are now in a position to extend the motivating example at the beginning of this subsection to the general case.
Agent $i$ can determine the monitoring result of $\varphi$ if, for each time $t \in [0, T + T_\varphi]$ and for each leaf node $q \in \{1,\dots, \beta\}$, at least one of the following conditions holds:
\begin{itemize}[leftmargin=*]
    \item 
    It holds that $\varphi_q^l = \top$ or $\varphi_q^l = \bot$, or agent $i$ has access to all the states of agents in $\hat{\mathcal{N}}_{q,t}(i)$, i.e., 
    \begin{equation}\label{eq:info1}
         (\varphi_q^l = \top \vee \varphi_q^l = \bot) \vee  (i.x_t^j \neq \bot, \text{for all} \ j \in \hat{\mathcal{N}}_{q,t}(i)).
    \end{equation}
    \item 
    The number of neighbors is less than the required minimum number, i.e., 
    \begin{align}\label{eq:info2}
        \Big|\Big\{i_1 \in \hat{\mathcal{N}}^{W_{p_1}, d_{p_1}}_{\mathcal{G}_{p_1}, t}(i) \ \Big| \ |\{i_2 \in \hat{\mathcal{N}}^{W_{p_2}, d_{p_2}}_{\mathcal{G}_{p_2}, t}(i_1) \mid \dots |\{i_{r} \in \hat{\mathcal{N}}^{W_{p_r}, d_{p_r}}_{\mathcal{G}_{p_r}, t}(i_{p_r-1}) \}| \ge E_{p_r}^{\min}  \dots\}| \ge E_{p_2}^{\min}\Big\}\Big| < E_{p_1}^{\min},
    \end{align}
where we recall that $a_q=(p_1, \dots, p_r)$ is the ancestor node list of the $q$th leaf node, and $E_{p_s}^{\min}$ is the minimum number in the counting interval $E_{p_s}$. 
\end{itemize}
We note that we have added ``$\varphi_q^l = \top$'' and ``$\varphi_q^l = \bot$'' in the first condition. This is because the states of agents in $\hat{\mathcal{N}}_{q,t}(i)$ are not required to determine the satisfaction of the \textsf{true} predicate $\top$ or the \textsf{false} predicate $\bot$. To help the reader understand the second condition, we briefly elaborate on equation \eqref{eq:info2}. If $a_q=(p_1)$, i.e., there is only one level, the condition \eqref{eq:info2} turns to be $|\{i_1 \in \hat{\mathcal{N}}^{W_{p_1}, d_{p_1}}_{\mathcal{G}_{p_1}, t}(i) \}| < E_{p_1}^{\min}$, meaning that the number of agent $i$'s neighbour has to be less than $E_{p_1}^{\min}$. If $a_q=(p_1, p_2)$, i.e., it there are two levels, the condition \eqref{eq:info2} turns to be $\Big|\{i_1 \in \hat{\mathcal{N}}^{W_{p_1}, d_{p_1}}_{\mathcal{G}_{p_1}, t}(i) \big| |\{i_2 \in \hat{\mathcal{N}}^{W_{p_2}, d_{p_2}}_{\mathcal{G}_{p_2}, t}(i_1)  \}| \ge E_{p_2}^{\min} \}\Big| < E_{p_1}^{\min}$, meaning that the number of agent $i$'s neighbors at level $p_1$ whose own number of neighbors at level $p_2$ is no less than $ E_{p_2}^{\min}$, is itself less than $ E_{p_1}^{\min}$.

Proposition \ref{pro:info} formally describes the sufficient condition for determining whether a distributed monitoring problem can yield a conclusive answer.

\begin{mypro}\label{pro:info}
    If, for all leaf nodes $q \in \{1,\dots, \beta\}$ and for all times $t \in [0, T+T_\varphi]$, either condition \eqref{eq:info1} or \eqref{eq:info2} is satisfied, then agent $i$ can determine that the value of $i.s_{\varphi, i}(t)$ is either 1 or 0.
\end{mypro}

\subsection{Distributed Monitoring of STL-GO-S}\label{subsec:dist_algo}

The distributed monitoring algorithm follows a similar approach as the centralized monitoring algorithm, with the aim of inductively computing the ternary signal $i.s_{\varphi, i}$ from the bottom up from the parse tree of the formula $\varphi$.
Recall that the first $i$ in $i.s_{\varphi, i}$ is the index of the agent's monitor. 
We will compute $i.s_{\varphi, i}$ by parsing all related subformulae $\varphi'$ in $\varphi $ and iterating over all relevant agents.
The main distinction lies in handling computations involving the unknown state $\bot$, and the signal state $?$.

The computation of the signal for the true predicate is the same as that of the centralized monitoring algorithm, i.e., $i.s_{\top, j}(t) = 1, \forall t \in [0, T+T_\phi]$ and $\forall j \in \mathcal{V}$.
We proceed with the computation of the ternary signal $i.s_{\varphi', j}$ for the atomic predicates $\varphi' \coloneqq \pi^{\mu_x}$, as described below,
\begin{align}
    i.s_{\varphi', j}(t) = 
    \begin{cases}
        \mathbf{1}(\mu_x(i.x^j_t) \geq 0)  & \text{if} \ i.x^j_t \neq \bot,  \\
        ? & \text{if} \ i.x^j_t = \bot.
    \end{cases} \nonumber
\end{align}

For the computation of signals corresponding to Boolean and temporal operators, we use the same equations as those in centralized monitoring while incorporating the truth table for three-valued logic \cite{malinowski1993many}. In particular, apart from the standard rules in binary logic, we account for the extra rules associated with three-valued logic: $1 - ? = ?$, i.e., $\neg ? = ?$; $0 \wedge ? = 0$; $1 \wedge ? = ?$; $0 \vee ? = ?$; $1 \vee ? = 1$.

For the computation of graph operators, $\varphi' \coloneqq \mathbf{In}^W_{\mathcal{G}, E} \varphi_1$ and $\varphi' \coloneqq \mathbf{Out}^W_{\mathcal{G}, E} \varphi_1$, we adopt a different approach from the centralized monitoring algorithm, which calculates the signal directly from the indicator function.
Instead, we define a signal list, $i.s_{\varphi_1, \hat{\mathcal{N}}^{W, d}_{\mathcal{G}, t}(j)}(t)$, representing the signals of agents in the neighbor set $\hat{\mathcal{N}}^{W, d}_{\mathcal{G}, t}(j)$ at time $t$, i.e., $i.s_{\varphi_1, \hat{\mathcal{N}}^{W, d}_{\mathcal{G}, t}(j)}(t) \coloneqq \{i.s_{\varphi_1, k}(t) \mid k \in \hat{\mathcal{N}}^{W, d}_{\mathcal{G}, t}(j)\}$.
We then define the number of satisfactions and non-violations as $n_{\textsf{sat}}(i.s_{\varphi_1, \hat{\mathcal{N}}^{W, d}_{\mathcal{G}, t}(j)}(t))$ and $n_{\neg \textsf{viol}}(i.s_{\varphi_1, \hat{\mathcal{N}}^{W, d}_{\mathcal{G}, t}(j)}(t))$, respectively, where these quantities represent the number of occurrences of ``1'' (satisfactions) and the combined occurrences of ``1'' and ``?'' (non-violations) in the signal list.
For simplicity, we denote $n_{\textsf{sat}}$ and $n_{\neg \textsf{viol}}$ to refer to $n_{\textsf{sat}}(i.s_{\varphi_1, \hat{\mathcal{N}}^{W, d}_{\mathcal{G}, t}(j)}(t))$ and $n_{\neg \textsf{viol}}(i.s_{\varphi_1, \hat{\mathcal{N}}^{W, d}_{\mathcal{G}, t}(j)}(t))$, respectively. 
Then, we have
\begin{align}\label{eq:dis}
    i.s_{\varphi', j}(t) =  
    \begin{cases}
        1 & \text{if} \ n_{\textsf{sat}} \ge e_1 \wedge n_{\neg \textsf{viol}} \le e_2, \\
        0 & \text{if} \ n_{\neg \textsf{viol}} < e_1 \vee  n_{\textsf{sat}} > e_2, \\
        ? & \text{otherwise},
    \end{cases} 
\end{align}
where $e_1$ and $e_2$ are minimum and maximum thresholds in the interval $E$, respectively.

We conclude this section that the monitor is sound: $s=1$ implies formula satisfaction, $s=0$ implies violation, and $s=?$ indicates uncertainty. Hence, the monitor solves Problem 2.

\section{Expressiveness of STL-GO}\label{sec:exp}

In this section, we will demonstrate the expressiveness of STL-GO by describing three important spatio-temporal properties in the MAS: counting, inter-agent distance, and agent-trace distance.
These properties were first formally described in SaSTL \cite{ma2020sastl}, SSTL\cite{nenzi2015specifying, nenzi2015qualitative}, and STREL \cite{bartocci2017monitoring, nenzi2022logic}.

\begin{figure}
    \centering
    \begin{tikzpicture}

    \filldraw[fill=gray, opacity=0.5] plot[smooth cycle] 
    (-0.5, 0) .. controls (-0.49, 0.2) and (-0.2, 0.49) .. (0, 0.5) 
        .. controls (1, 0.5) and (2, 0.3) .. (2.5, 0.7) 
        .. controls (3, 1) and (3.7, 2.05) .. (4.2, 2.06) 
        .. controls (4.5, 2.05) and (5.3, 1.2) .. (5.35, 0.6) 
        .. controls (5.35, 0) and (4, -0.4) .. (3, -0.5) 
        .. controls (2, -0.5) and (1, -0.5) .. (0, -0.5) 
        .. controls (-0.2, -0.49) and (-0.49, -0.2) .. (-0.5, 0);
    
    \node[style={circle, draw, thick, minimum size=6mm} ](1) at (0,0) {1};
    \node[style={circle, draw, thick, minimum size=6mm} ] (3) [above right of=1, xshift=0.8cm, yshift=0.4cm] {3};
    \node[style={circle, draw, thick, minimum size=6mm} ] (2) [above left of=3, xshift=-0.8cm, yshift=0.4cm] {2};
    \node[style={circle, draw, thick, minimum size=6mm} ] (4) [below right of=3, xshift=0.8cm, yshift=-0.4cm] {4};
    \node[style={circle, draw, thick, minimum size=6mm} ] (5) [above right of=3, xshift=0.8cm, yshift=0.4cm] {5};
    \node[style={circle, draw, thick, minimum size=6mm} ] (6) [above right of=4, xshift=1.2cm, yshift=-0.1cm] {6};
    \node[style={circle, draw, thick, minimum size=6mm} ] (7) [above left of=6, yshift=0.3cm] {7};
    
    \draw (1) -- (3);
    \draw (2) -- (3);
    \draw (3) -- (4);
    \draw (3) -- (5);
    \draw (4) -- (6);
    \draw (5) -- (7);
    \draw (6) -- (7);

    \draw[dashed, red, thick] (4.6, 1.15) ellipse (0.9cm and 1cm);
    \draw[thick] (5.1, 2) -- (5.5, 2.5);
    \node[draw=none] at (5.5, 2.75) {\small{information center}};
    \node[draw=none] at (1.5, -0.3) {\small{signal coverage area}};

    \node[draw=none] at (0, 0.45) {\small{\textsf{H}}};
    \node[draw=none] at (3, 0.45) {\small{\textsf{H}}};
    \node[draw=none] at (0, 2.7) {\small{\textsf{H}}};
    \node[draw=none] at (3, 2.7) {\small{\textsf{H}}};
    \node[draw=none] at (4.2, 2.1) {\small{\textsf{H}}};
    \node[draw=none] at (4.9, 1.1) {\small{\textsf{H}}};
    \node[draw=none] at (1.5, 1.55) {\small{\textsf{C}}};

    \node[draw=none] at (0.65, 0.7) {\small{6}};
    \node[draw=none] at (0.65, 1.9) {\small{8}};
    \node[draw=none] at (2.3, 0.7) {\small{8}};
    \node[draw=none] at (2.3, 1.9) {\small{6}};
    \node[draw=none] at (3.8, 0.4) {\small{8}};
    \node[draw=none] at (3.7, 2) {\small{8}};
    \node[draw=none] at (4.65, 1.15) {\small{8}};
    
    \end{tikzpicture}
    \caption{An example of distance graph.}
    
    \label{fig:dist}
\end{figure}
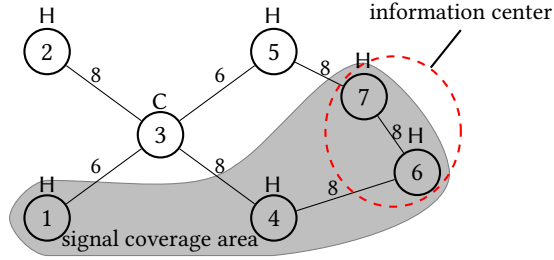

Counting the number of agents satisfying a spatio-temporal property is a common task in multi-agent systems, as it provides a measure of the system’s overall performance and behavior. 
For example, in a logistics warehouse, there are two types of robots: coordinator robots and handling robots, denoted by atomic propositions \textsf{C} and \textsf{H}, respectively.
The coordinator robot manages and coordinates a specific area of handling robots, and the handling robots are responsible for transporting goods. 
At time $t$, the distance graph between these robots is shown in Fig.~\ref{fig:dist}, where agents' atomic propositions and weights of distance are marked near nodes and edges, respectively.
We may be interested in tasks such as, \emph{at all times, at least 2 handling robots within 10 meters of the coordinator robot should stay in the signal coverage area (marked as the grey area in Fig.~\ref{fig:dist})}. In this case, we count the number of agents satisfying the atomic propositions \textsf{H} and spatial requirement.

Inter-agent and agent-trace distances are two different types of information about distances in a multi-agent system.
Specifically, agents can be indirectly connected through other intermediate agents. Information about this series of agents forming a connection is referred to as “agent-trace” information. On the other hand, when we only focus on the shortest distance of the two endpoints of this series over the graph, we refer to it as “inter-agent” information. In contrast, agent-trace information considers the entire trace of agents connecting the endpoints, including the states and distances of all agents along the path.
We may be interested in inter-agent distance properties such as, \emph{at all times, there exists a handling robot within 10 meters of the coordinator robot that is in the signal coverage area in a shortest distance sense on the graph}.
In this task, we only consider one handling robot and one coordinator robot, which are two endpoints.
Additionally, we may be interested in the agent-trace distance properties such that \emph{at time $t$, an agent in the information center (marked as the ellipse with red line in Fig.~\ref{fig:dist}) that can be reached from the coordinator robot through any trace of other agents. The length of the trace should be less than 20 meters, and intermediate agents should stay in the signal coverage area, to make the information delivered to the agent in the information center}.

In the following three subsections, we will demonstrate how STL-GO can capture these properties by presenting their equivalent formulations with our graph-based operators, aligning them with the operators introduced in SaSTL \cite{ma2020sastl}, SSTL\cite{nenzi2015specifying, nenzi2015qualitative}, and STREL \cite{bartocci2017monitoring}.

The comparison is summarized in Table \ref{table:1}, where counting, inter-agent distance, and agent-trace distances are discussed in the following subsections. Note that  \emph{multi-graph info} characterizes whether a formula is capable of expressing the task of a MAS evolving over multiple graphs, which can be only described by our proposed STL-GO-S and STL-GO. Furthermore, only the component $i.\varphi$ formulation in STL-GO enables task specifications to be imposed on multiple agents.

\begin{table}[htb] 
    \begin{center}    
    \caption{Expressiveness comparison.}
    \label{table:1}  
    \begin{threeparttable}[b]
        \begin{tabular}{ccccccccc}
            \toprule
            \small{\textbf{Tasks}} & \small{STL-GO} & \small{STL-GO-S} & \small{SaSTL \cite{ma2020sastl}} & \small{SSTL\cite{nenzi2015specifying, nenzi2015qualitative}} & \small{STREL \cite{bartocci2017monitoring, nenzi2022logic}} \\
            \midrule
            \small{Counting} & \checkmark & \checkmark & \checkmark & \ding{55} & \ding{55}  \\
		\small{Inter-agent distance} & \checkmark & \checkmark & \checkmark & \checkmark  & \checkmark   \\
		\small{Agent-trace distance} & \checkmark & \checkmark & \ding{55} & \ding{55} & \checkmark   \\
        \small{Multi-graph info} & \checkmark & \checkmark & \ding{55} & \ding{55}  & \ding{55}  \\
        \small{Imposed on multi-agents} & \checkmark & \ding{55} & \ding{55} & \ding{55} & \ding{55}   \\
            \bottomrule
        \end{tabular}
    \end{threeparttable}
    \end{center}   
\end{table}

\subsection{Counting}
\label{sec:counting}
To count the number of neighboring agents satisfying spatio-temporal properties, SaSTL \cite{ma2020sastl} introduces a “Counting” operator that quantifies the number of agents meeting a specified condition.
In this subsection, we show the ``Counting'' operator in \cite{ma2020sastl} can be equivalently represented using graph operators and comment on the differences and similarities between our logic and SaSTL. 

The ``Counting'' operator is based on the shortest distance graph $\mathcal{G}^{d_s} \coloneqq (\mathcal{V}, \mathcal{E}^{d_s}, w^{d_s})$ constructed over a distance graph $\mathcal{G}^d$.
Its syntax is given by $\mathcal{C}_{(W, \psi)}^{op} \varphi \sim c$, where $W$ is the distance interval, $op \in \{sum, avg\}$ \footnote{Note that operators ``max'' and ``min'' are excluded here, as \cite{ma2020sastl} primarily utilizes them to define the “Somewhere” and “Everywhere” operators, which will be discussed in the next subsection.}  is the operator, and $c$ is an integer. Here,
$\psi$ is a label indicating the agent’s property over a set of atomic propositions, such as \textsf{C} and \textsf{H} in the example at the beginning of this section.
The semantics of the Counting operator with $op = sum$ is defined as $(\mathcal{MA}, i, t) \models \mathcal{C}_{(W, \psi)}^{sum} \varphi \sim c$ iff $|\{j \in \mathcal{V} \mid w^{d_s}_t(i, j) \in W \wedge (\mathcal{MA}, j, t) \models \varphi \wedge j \models \psi\}| \sim c$, where $j \models \psi$ indicates that agent $j$ possesses the label $\psi$, and $\sim \in \{\le, <\}$. 

In STL-GO, we can describe ``counting'' with the sum operator using graph operators $\mathbf{In}^W_{\mathcal{G}, E}$ (or $\mathbf{Out}^W_{\mathcal{G}, E}$) by a new type of graph, called the $\psi$-labeled graph with agent $i$. The $\psi$-labeled distance graph with agent $i$ is defined as $\mathcal{G}^{\psi,i} \coloneqq (V^{\psi, i}, \mathcal{E}^{\psi, i}, w^{\psi, i})$, where $V^{\psi, i} \coloneqq \{j \in \mathcal{V} \mid j \models \psi\} \cup \{i\}$,
$(j, k) \in \mathcal{E}^{\psi, i}$ iff $(j, k) \in \mathcal{E}^{d_s}$, and $w^{\psi,i}(j, k) = w^{d_s}(j, k)$.
Intuitively, $\mathcal{G}^{\psi,i}$ is a sub-graph of  graph $\mathcal{G}^{d_s}$, where nodes contain agent $i$ and agents satisfying property $\psi$.

\begin{remark}
    It is worth mentioning that constructing custom graphs, such as the $\psi$-labeled distance graph $\mathcal{G}^{\psi,i}$, can be fully automated. In practice, users are not required to manually create these graphs, as they can be generated algorithmically prior to runtime. 
    Furthermore, such graphs can be precomputed and included in the predefined graph set $\mathcal{T}$ used for task specification and evaluation. For example, in scenarios where both a nominal distance graph and a shortest distance graph are required—such as when a task specifies that the shortest distance to a target should be less than 100 meters, while also requiring the existence of two nominal trajectories that reach the destination within the same distance—both graphs can be included in $\mathcal{T}$ for simultaneous use.
    Even if the shortest distance graph is not initially available, it can be automatically constructed from the nominal distance graph using standard graph algorithms (e.g., Dijkstra’s algorithm). This flexibility allows $\mathcal{T}$ to include all necessary graphs for complex multi-agent task formulations without additional manual effort from the user.
\end{remark}

Then, we can equivalently express $\mathcal{C}_{(W, \psi)}^{sum} \varphi \sim c $ using graph $\mathcal{G}^{\psi,i}$ as follows.

\begin{mypro}
    Let $\mathcal{C}_{(W, \psi)}^{sum} \varphi \sim c$ be an SaSTL formula defined in \cite{ma2020sastl}.
    Given a labeled distance graph $\mathcal{G}^{\psi,i}$ with agent $i$, we have 
    \begin{align}
        (\mathcal{MA}, i, t) \models \mathcal{C}_{(W, \psi)}^{sum} \varphi \sim c  \ \text{iff} \ (\mathcal{MA}, t) \models i.(\mathbf{In}_{\{\mathcal{G}^{\psi, i}\}, C'}^{W} \varphi), \nonumber 
    \end{align}
    where $C' \coloneqq \{c' \in \mathbb{N} \cup \{+\infty\} \mid c' \sim c \}$ is the interval of the counting number. 
    We can also write it by STL-GO-S as $(\mathcal{MA}, i, t) \models \mathcal{C}_{(W, \psi)}^{sum} \varphi \sim c  \ \text{iff} \ (\mathcal{MA}, i, t) \models \mathbf{In}_{\mathcal{G}^{\psi, i}, C'}^{W} \varphi$.
\end{mypro}

We acknowledge that a general equivalent description of the $avg$ operation is not feasible with our logic, as the exact total number of neighboring agents satisfying label requirement $\psi$ and the distance requirement is not computable via our logic. However, suppose $N'$ denotes the total number of neighboring agents satisfying $\psi$ and the distance requirement. Suppose further that $N'$ is known beforehand. Then $(\mathbf{x}, i, t) \models \mathcal{C}_{(W, \psi)}^{avg} \varphi \sim c$ iff $(\mathcal{MA}, t) \models i.(\mathbf{In}_{\{\mathcal{G}^{\psi, i}\}, C'}^{W} \varphi)$, where $C' \coloneqq \{c' \in \mathbb{N} \cup \{+\infty\} \mid c' \sim c \times N' \}$. In practice, for centralized monitoring, this is not a limiting assumption since one can always compute $N'$ in advance under Assumption 1 and when the MAS states are known (which is also an assumption in SaSTL). Observing the similarity between the counting operator and In/Out operator, we emphasize on our strength (practically motivated by our examples) in achieving the following goals with succinct formulae, which are not explicitly considered in SaSTL: a) we reason over multiple graph topologies with the graph quantifiers $\exists$ and $\forall$, b) we consider possibly directed multigraphs where multiple edges connect two agents, c) we consider STL-GO on top of STL-GO-S as a logic defined globally over the multiagent system, and d) we allow nested counting properties where we count the neighbors of a node where the neighbors also satisfy counting specifications. We also remark on our drawback as compared to SaSTL \cite{ma2020sastl} in that we are generally unable to express counting with the average operator and aggregations (which measure the aggregated property over neighboring states).

Consider the example at the beginning of this section.
For SaSTL, we can describe it as $\mathbf{G}_{[0,\infty]} \mathcal{C}^{sum}_{([0, 10], H)}\pi^{\mu_1} \ge 2$, where $\mu_1$ is the predicate function for signal coverage area.
We can also write it by STL-GO-S $\varphi = \mathbf{G}_{[0,\infty]} \mathbf{In}^{[0, 10]}_{\{\mathcal{G}^{\textsf{H}, 3}\}, [2, \infty]} \pi^{\mu_1}$ and imposing it on agent 3, which is the coordinator agent, or by STL-GO $\phi = 3.\mathbf{G}_{[0,\infty]} \mathbf{In}^{[0, 10]}_{\{\mathcal{G}^{\textsf{H}, 3}\}, [2, \infty]} \pi^{\mu_1}$.


\subsection{Inter-agent distance} \label{subsec:behaviorofdistance_graph}

SSTL \cite{nenzi2015specifying, nenzi2015qualitative} proposed the operators ``Somewhere'' and ``Everywhere'' to capture inter-agent distance information. 
In this subsection, we show the equivalence between the ``Somewhere'' and ``Everywhere'' operators in SSTL and our graph operators.

``Somewhere'' and ``Everywhere'' are both defined over the shortest distance graph.
Their syntax is $\somewhere_{W} \varphi$ and $\everywhere_{W} \varphi$, where $W$ is the interval of the distance. 
The semantics of ``Somewhere'' is defined as $(\mathcal{MA}, i, t) \models \somewhere_{W} \varphi$ iff there exists $j \in \mathcal{V}$ such that $w_t^{d_s}(i, j) \in W$ and $(\mathcal{MA}, j, t) \models \varphi$. 
``Everywhere'' can be derived from ``Somewhere'' by $\everywhere_{W} \varphi \coloneqq \neg \somewhere_{W} \neg \varphi$, and its semantics is defined as $(\mathcal{MA}, i, t) \models \everywhere_{W} \varphi$ iff for all $j \in \mathcal{V}$ such that $w_t^{d_s}(i, j) \in W$ and $(\mathcal{MA}, j, t) \models \varphi$.
Note that ``Somewhere'' and ``Everywhere'' operators are introduced based on the shortest distance graph.

Then, we can equivalently express the ``Somewhere'' and ``Everywhere'' operators using the shortest distance graph $\mathcal{G}^{d_s}$ by STL-GO  as follows.

\begin{mypro}
    Let $\somewhere_{W} \varphi$ and $\everywhere_{W} \varphi$ be SSTL formulae defined in \cite{nenzi2015specifying, nenzi2015qualitative}.
    Given a shortest distance graph $\mathcal{G}^{d_s}$, we have 
    \begin{align}
        (\mathcal{MA}, i, t) \models \somewhere_{W} \varphi  \ \text{iff} \ (\mathcal{MA}, t) \models i.(\mathbf{In}_{\{\mathcal{G}^{d_s}\}, [1, +\infty]}^{W} \varphi), \nonumber \\
        (\mathcal{MA}, i, t) \models \everywhere_{W} \varphi  \ \text{iff} \ (\mathcal{MA}, t) \models i.(\mathbf{In}_{\{\mathcal{G}^{d_s}\}, [0, 0]}^{W} \neg \varphi). \nonumber
    \end{align}
    We can also write them by STL-GO-S as $(\mathcal{MA}, i, t) \models \somewhere_{W} \varphi  \ \text{iff} \ (\mathcal{MA}, i, t) \models \mathbf{In}_{\mathcal{G}^{d_s}, [1, +\infty]}^{W} \varphi$ and $(\mathcal{MA}, i, t) \models \everywhere_{W} \varphi  \ \text{iff} \ (\mathcal{MA}, i, t) \models \mathbf{In}_{\mathcal{G}^{d_s}, [0, 0]}^{W} \neg \varphi$.
\end{mypro}


Consider the example at the beginning of this section.
For SSTL, we can describe it as $\mathbf{G}_{[0,\infty]} \somewhere_{[0,10]} \pi^{\mu_1}$, where $\mu_1$ is the predicate function for signal coverage area.
We can also write it by STL-GO-S $\varphi = \mathbf{G}_{[0,\infty]} \mathbf{In}^{[0, 10]}_{\{\mathcal{G}^{d_s}\}, [1, \infty]} \pi^{\mu_1}$ and imposing it on agent 3, or by STL-GO $\phi = 3.\mathbf{G}_{[0,\infty]} \mathbf{In}^{[0, 10]}_{\{\mathcal{G}^{d_s}\}, [1, \infty]} \pi^{\mu_1}$.

\subsection{Agent-trace distance}\label{subsec:behaviorofdistance_trace}

STREL \cite{bartocci2017monitoring, nenzi2022logic} proposed operators to capture the agent-trace distance information.
In this subsection, we show the equivalent relation between the ``Reach'' and ``Escape'' operators in STREL and our graph operators.

Before presenting their syntax and semantics, we first introduce a definition called ``agent-trace''. A trace, denoted as $\tau \coloneqq l_0 \dots l_{|\tau|-1}$, is a sequence, where $\forall i \in \{0, \dots, |\tau|-1\}: l_i \in \mathcal{V}$ and $\forall i \in \{0, \dots, |\tau|-2\}$: $(l_i, l_{i+1}) \in \mathcal{E}^d$, defined in section \ref{sec:counting}. Here $|\tau|$ is the length of the trace and $l_i$ is the index of the agents appearing in the $i+1$th of the trace.

The syntax of ``Reach'' and ``Escape'' are $\varphi_1 \mathbf{R}_w^f \varphi_2$ and $ \mathbf{E}_w^f \varphi$, where they define $w$ as a weight predicate, and $f$ is the distance function.
In order to keep the same setting as in our paper, we simplify the distance function $f$ to be the sum of weights along the agent-trace and distance predicate $w$ to be $w \in W \coloneqq [w_1, w_2]$. 
Then, we can write $\varphi_1 \mathbf{R}_w^f \varphi_2$ and $\mathbf{E}_w^f \varphi$ as $\varphi_1 \mathbf{R}_{W} \varphi_2$ and $\mathbf{E}_{W} \varphi$.
The semantics of the ``Reach'' operator is defined as $(\mathcal{MA}, i, t) \models \varphi_1 \mathcal{R}_{W} \varphi_2$ iff there exists an agent-trace $\tau \coloneqq l_0 l_1 \dots l_{|\tau|-1}$ with $l_0 = i$ such that 
(1) $(\mathcal{MA}, j, t) \models \varphi_1, \forall j \in \{l_0, l_1, \dots l_{|\tau|-2}\}$; 
(2) $(\mathcal{MA}, l_{|\tau|-1}, t) \models \varphi_2$; and 
(3) $\sum_{j=0}^{|\tau|-2}w^d_t((l_j, l_{j+1})) \in W$.
The semantics of the ``Escape'' operator is defined as $(\mathcal{MA}, i, t) \models \mathbf{E}_{W} \varphi$ iff there exists an agent-trace $\tau \coloneqq l_0 l_1 \dots l_{|\tau|-1}$ with $l_0 = i$ such that  
(1) $(\mathcal{MA}, j, t) \models \varphi, \forall j \in \{l_0, l_1, \dots l_{|\tau|-1}\}$;  and 
(2) $w_t^{d_s}((l_0, l_{|\tau|-1})) \in W$, where $w_t^{d_s}((l_0, l_{|\tau|-1}))$ is the shortest distance between $l_0$ and $l_{|\tau|-1}$ in the shortest distance graph $\mathcal{G}^{d_s}_t$.
Intuitively, the reachability operator $\varphi_1 \mathbf{R}_{W} \varphi_2$ describes the behavior of reaching an agent satisfying property $\varphi_2$, through a path with all agents that satisfy $\varphi_1$, and with a distance that belongs to $W$, while the escape operator $\mathbf{E}_{W} \varphi$ describes the possibility of escaping from a certain region via a route passing only through locations that satisfy $\varphi$, with the distance between the starting location of the path and the last that belongs to $W$. 
The main difference between these two operators is that the distance of the reach operator is with respect to the path, instead, the distance of the escape operator is between agents, so it considers the shortest path distance between the starting agent and the last.

In STL-GO, we can describe the ``Reach'' and ``Escape'' operators by introducing the sets of traces.
Specifically, we construct two trace sets $\mathsf{Trace}_{W, t}^{i, \mathbf{R}}$ and $\mathsf{Trace}_{W, t}^{i,\mathbf{E}}$, which contain all the traces that we will consider in operators ``Reach'' and ``Escape''. A trace $\tau \coloneqq l_0 \dots l_{|\tau|-1} \in \mathsf{Trace}_{W, t}^{i, \mathbf{R}}$ if $l_0 = i$ and $\sum_{j=0}^{|\tau|-2}w^d_t((l_j, l_{j+1})) \in W$, and a trace $\tau \coloneqq l_0 \dots l_{|\tau|-1} \in \mathsf{Trace}_{W, t}^{i,\mathbf{E}}$ if $l_0 = i$ and $w^{d_s}_t((l_0, l_{|\tau|-1})) \in W$.
Then, we can equivalently express the ``Reach'' and ``Escape'' operators using the sets of traces $\mathsf{Trace}_{W, t}^{i, \mathbf{R}}$ and $\mathsf{Trace}_{W, t}^{i,\mathbf{E}}$ as follows.

\begin{mypro}
    Let $\varphi_1 \mathbf{R}_{W} \varphi_2$ and $\mathbf{E}_{W} \varphi$ be STREL formulae defined in \cite{bartocci2017monitoring}. Given two trace sets $\mathsf{Trace}_{W, t}^{i, \mathbf{R}}$ and $\mathsf{Trace}_{W, t}^{i,\mathbf{E}}$, we have
    \begin{align}
        & (\mathcal{MA}, i, t)  \models \varphi_1 \mathbf{R}_{W} \varphi_2 \ \text{iff} \ \
        (\mathcal{MA}, t) \models \vee_{\tau \in \mathsf{Trace}_{W, t}^{i, \mathbf{R}}} (\mathbf{FA}_{\tau[:-1]} \varphi_1 \wedge \tau[-1].\varphi_2), 
        \nonumber   \\
        & (\mathcal{MA}, i, t)  \models \mathbf{E}_{W} \varphi \ \text{iff} \ \ (\mathcal{MA}, t) \models \vee_{\tau \in \mathsf{Trace}_{W, t}^{i, \mathbf{E}}} (\mathbf{FA}_{\tau[:]} \varphi). \nonumber
    \end{align}
    where 
    $\tau_l[:-1] \coloneqq \{l_0, \dots, l_{|\tau|-2}\}$ is the set of agents' indices in the trace except the last agent, $\tau_l[-1] = l_{|\tau|-1}$ is the index of the last agent in the trace, and $\tau[:]\coloneqq \{l_0, \dots, l_{|\tau|-1}\}$ is the set of all agents' indices in the trace.
\end{mypro}

Consider the example at the beginning of this section. For STREL, we can describe it as $\pi^{\mu_1} \mathbf{R}_{[0,20]} \pi^{\mu_2} $, where $\mu_1$ and $\mu_2$ are the predicate functions for the signal coverage area and the information center, respectively.
We can equivalently write it by STL-GO $\phi = \vee_{\tau \in \mathsf{Trace}_{[0,20], t}^{3, \mathbf{R}}} (\mathbf{FA}_{\tau[:-1]} \pi^{\mu_1} \wedge \tau[-1].\pi^{\mu_2})$, where $\mathsf{Trace}_{[0,20], t}^{3, \mathbf{R}} = \{\tau_1, \dots, \tau_6\}$ with $\tau_1 = 1$, $\tau_2 = 2$, $\tau_3 = 4$, $\tau_4 = 5$, $\tau_5 = 5, 7$, and $\tau_6 = 4,6$.

\begin{remark}
    If the distance function $f$ is defined as the sum of the hops, or all the weights in the graph are 1, then, we can also describe the ``Reach'' and ``Escape'' operators by nesting the graph operator $\mathbf{In}^W_{\mathcal{G}, C}$.
    For example, $(\mathcal{MA}, i, t)  \models \varphi_1 \mathbf{R}_{[1,2]} \varphi_2$ iff $(\mathcal{MA}, t) \models i.\big((\varphi_1 \wedge \mathbf{In}_{\{\mathcal{G}^d\}, [1, +\infty]} (\varphi_1 \wedge \mathbf{In}_{\{\mathcal{G}^d\}, [1, +\infty]} \varphi_2)) \vee (\varphi_1 \wedge \mathbf{In}_{\{\mathcal{G}^d\}, [1, +\infty]} \varphi_2)\big)$.
\end{remark}

\section{Examples}\label{sec:example}
In this section, we present two examples to illustrate the expressiveness of STL-GO in multi-agent systems, and we empirically validate the centralized and the distributed monitoring algorithms.

\subsection{Bike-Sharing System}
We apply STL-GO to monitor the Jersey City bike-sharing system using publicly available data from the Citi Bike platform \cite{bikewebsite}. Our analysis covers the period from October 1 to October 31, 2024.
Specifically, there are 118 stations in Jersey City, with each station represented as an agent in a multi-agent system, i.e., $\mathcal{V} = \{1, \dots, 118\}$. 
The sampling time of this system is one hour, and the task specifications are based on a 24-hour (one-day) period.
The state of station $i$ at time $t$ is modeled as $x^i_t \coloneqq [n^i_t, n^{i,in}_t, n^{i,out}_t]$ with dynamics $n^i_{t+1} = n^i_t + n^{i,in}_t - n^{i,out}_t$ for all $t \in \{0, \dots, 24\}$, where $n^i_t$, $n^{in,i}_t$, and  $n^{out,i}_t$ are the total number of bikes, number of incoming bikes, and number of outgoing bikes at station $i$ at time $t$, respectively.
We construct a  single-edge distance graph $\mathcal{G}^d$ with a distance function $w^d(i, j)$ representing the biking distance from station $i$ to station $j$, where the distance is collected from the Google Maps. 
We also construct a multigraph, denoted by $\mathcal{G}^{mt}$, to represent both the public transportation time and walking time between two stations, where $mt$ stands for multiple time.
In this multigraph, $w^{mt}(i,j,1)$ and $w^{mt}(i,j,2)$ represent the public transportation time and walking time from station $i$ to station $j$, respectively.
Both graphs are directed since the biking distance from station $i$ to station $j$ may be different to the biking distance from station $j$ to station $i$, e.g., due to the existence of one-way streets.
We assume that these graphs are time-invariant, which is often a reasonable assumption, e.g., public transportation times are mostly constant and biking times between two stations remains approximately the same. In the remainder, we describe various specifications for this bike-sharing system via STL-GO-S and STL-GO formulas. We also show the results of our centralized and distributed monitors.


One of the most popular stations is ``Grove St PATH'', and we are interested in monitoring the bike availability and analyzing the bike capacities, such as insufficiently low or high number of bikes. Specifically, we may be interested in the specification ``if the bike availability at “Grove St PATH” drops below 5, there are at least 5 strategies (edges) to arrive at another station within 8 minutes distance that has at least 8 bikes available'' to allow users to instead use a nearby station to find a ride. This specification can be expressed by STL-GO-S as follows:
\begin{align}
    \varphi_1 \coloneqq & \mathbf{G}_{[0, 24]} \big( n  <  5 \to \mathbf{Out}^{[0, 8]}_{\{\mathcal{G}^{mt}\}, [5, +\infty]} n \ge 8 \big). \nonumber
\end{align}
Additionally, when more than 15 bikes arrive at the station, we examine whether the net increase in bikes at up to 4 nearby stations (reachable from ``Grove St PATH'' within a 2-mile walking distance) are more than 5, to prevent bikes from clustering in certain areas:
\begin{align}
    \varphi_2 \coloneqq & \mathbf{G}_{[0, 24]} \big( n^{in} > 15 \to \mathbf{In}^{[0, 2]}_{\{\mathcal{G}^d\}, [0, 4]} n^{in} -n^{out} > 5 \big). \nonumber
\end{align}

Beyond individual station monitoring, we are interested in the overall performance of the bike-sharing system. 
To achieve this, we randomly sample 30 stations located within the city’s core, and denote this set as $V$. For each station in $V$, we require that there are at least 3 nearby stations within a 1-mile walking distance, each with at least 8 bikes available, which can be described by the following STL-GO formula:
\begin{align}
    \phi_1 \coloneqq & \mathbf{FA}_{V}\mathbf{G}_{[0, 24]}  \big( \mathbf{Out}^{[0, 1]}_{\{\mathcal{G}^d\}, [3, +\infty]} n \ge 8 \big). \nonumber
\end{align}
This requirement helps ensure a well-balanced system that enhances user experience.
The property of $\varphi_1$ can also be extended to all stations in set $V$, but with relaxed thresholds since most stations are not as heavily utilized as “Grove St PATH”:
\begin{align}
    \phi_2 \coloneqq & \mathbf{FA}_{V}\mathbf{G}_{[0, 24]}  \big( n<2 \to \mathbf{Out}^{[0, 12]}_{\{\mathcal{G}^{mt}\}, [3, +\infty]} n \ge 4 \big). \nonumber
\end{align}

We use the centralized monitoring algorithm in Section \ref{subsec:centralized} to repeatedly monitor the four aforementioned tasks over 31 days in October 2024. We present the results of centralized monitoring in Table \ref{table:case_1_centralized} where $\#$ sat denote the number of satisfactions and avg time denotes the average monitoring time in seconds across the 31 days. We also apply the distributed algorithm in Section \ref{sec:distributed} to monitor $\varphi_1$ and $\varphi_2$ again for the station "Grove St PATH". We assume it has access to data of all stations within weights 2.5 miles in $\mathcal{G}^d$, and that of stations within weights 7 minutes $\mathcal{G}^{mt}$. We present the results of distributed monitoring in Table \ref{table:case_1_distributed} where $\#$ sat, $\#$ vio, and $\#$ unknown denote the number of satisfactions, violations, and unknowns in the monitoring results across the 31 days respectively. In Table \ref{table:case_1_distributed}, avg time denotes the average monitoring time across the 31 days for the distributed monitoring algorithm.

\begin{table}[htbp]
  \centering
  \renewcommand{\arraystretch}{1.2}
  \begin{tabular}{|l|c|c|}
    \hline
    & \# sat & avg time (s) \\ \hline
    $\varphi_{1}$ & 31  &  $1.10 \times 10^{-4}$\\ 
    $\varphi_{2}$  &12  &  $3.36 \times 10^{-4}$\\ 
    $\phi_{1}$ &29  &  $2.13 \times 10^{-3}$\\
    $\phi_{2}$ &19 &  $2.73 \times 10^{-3}$\\ \hline
  \end{tabular}
  \caption{Centralized monitoring results.\label{table:case_1_centralized}}
\end{table}
\begin{table}[htbp]
  \centering
  \renewcommand{\arraystretch}{1.2}
 \begin{tabular}{|l|c|c|c|c|}
    \hline
    & \# sat & \# vio & \# unknown & avg time (s) \\ \hline
    $\varphi_{1}$ & 30& 0 & 1 &  $7.48 \times 10^{-5}$\\ 
    $\varphi_{2}$ & 12   & 0 & 19 & $6.33 \times 10^{-4}$ \\ \hline
  \end{tabular}
  \caption{Distributed monitoring results.\label{table:case_1_distributed}}
\end{table}


\subsection{Drone Surveillance}\label{subsec:drone}
\begin{figure}
    \centering
    \includegraphics[width =0.8\linewidth]{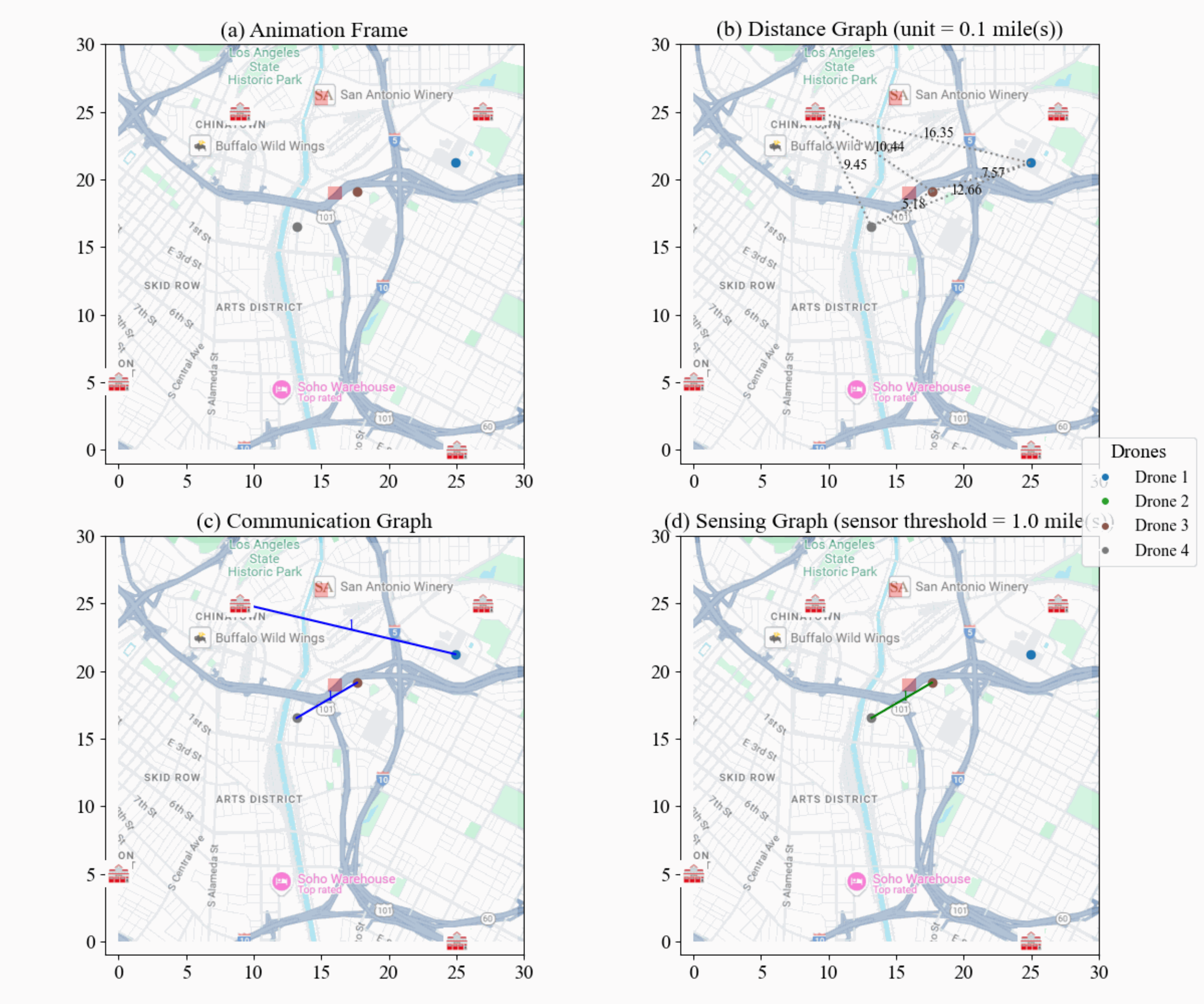}
    \caption{Example frame: drone surveillance simulation.}
    \label{fig:fire}
\end{figure}

We now apply our monitoring algorithms to a swarm of drones that are required to surveil various regions of interest. We consider a synthetic setting where $\sigma \in \mathbb{N}$ drone stations possess one drone each. We treat each drone as an agent (i.e., $\mathcal{V} = \{1, \hdots, \sigma\}$) and denote their locations, respectively, with $x^1, \hdots, x^\sigma$, where $x^i_t \in \mathbb{R}^2$ denotes the position of drone $i \in \{1,\hdots, \sigma\}$ at time $t$. For illustration, we simulate situations on the ground that are to be observed by drones, i.e., we want to dispatch drones within the city to perform surveillance tasks. We create a scenario of the simulation environment which is illustrated in Figure \ref{fig:fire}, where $\sigma = 4$. Specifically, the locations of the stations are shown with the house symbol, while drone locations are shown with dots in respective colors. We represent regions of interests with red squares in Figure \ref{fig:fire} (a).

In our simulation, when we require a drone to surveil a region of interest, we dispatch an available drone from any of the four stations to the region of interest. After the region is investigated, the drone must report back to its station before being dispatched to a new region. We assume drones $i$ where $i \le \lfloor\sigma / 2\rfloor$ belong to a specific category of drones, while all other drones to another category. Drones within the same category travel with the same speed. We represent the distance graph $\mathcal{G}^d$ as a complete undirected graph where $w^d(i, j) \coloneq \|x^i - x^j\|_2$, and we show the distance graph corresponding to Figure \ref{fig:fire} (a) in Figure \ref{fig:fire} (b). We allow drones with the same category to be always able to communicate with each other and thus model the communication graph $\mathcal{G}^c$ as a time-invariant undirected graph where $w^c(i, j) \coloneq 1$ if drone $i$ and drone $j$ are with the same category and make nodes $i$ and $j$ disconnected if the drones do not belong to the same category. We illustrate $\mathcal{G}^c$ corresponding to Figure \ref{fig:fire} (a) in Figure \ref{fig:fire} (c). To model the sensing topology, we allow two drones within the same category to be able to sense each other whenever they are close (the distance between the two drones are within $1$ mile). Formally, we define $\mathcal{G}^s$ as the sensing graph where nodes $i$ and $j$ are connected with weight $w^s(i, j) \coloneq 1$ if and only if $\|x^i - x^j\|_2 \le 1$ and drone $i$ and $j$ are with the same category. We illustrate $\mathcal{G}^s$ corresponding to Figure  \ref{fig:fire} (d).

We are interested in monitoring the safety of a drone in that it maintains a safe distance (at least 0.3 miles) from all other drones with a monitoring horizon of $2$ minutes into the future. This can be represented with an STL-GO-S formula $\varphi_3$ where
\begin{align}
    \varphi_3 \coloneqq & \mathbf{G}_{[0, 2]} (\mathbf{Out}^{[0.3, \infty]}_{\{\mathcal{G}^d\}, \{\sigma - 1\}}\top). \nonumber
\end{align}
We are also interested in monitoring if a drone can both sense and communicate with another drone eventually within $2$ minutes, which is represented with the following STL-GO-S formula
\begin{align}
    \varphi_4 \coloneqq \mathbf{F}_{[0, 2]} (\mathbf{Out}^\forall_{\{\mathcal{G}^s, \mathcal{G}^c\}, [1, \sigma - 1]} \top). \nonumber
\end{align}
Apart from monitoring the status of each drone separately, we also want to monitor if all drones maintain a safe distance from each other. We thus investigate the following STL-GO formula
\begin{align}
    \phi_3 \coloneqq \mathbf{G}_{[0, 2]} \textbf{FA}_{\mathcal{V}} (\mathbf{Out}^{[0.3, \infty]}_{\{\mathcal{G}^d\}, \{\sigma - 1\}} \top). \nonumber
\end{align}   
Lastly, we want to ensure that drones close to drone $i$ (within at most 1 mile), where $i$ is a pre-selected node, can sense or communicate with drone $i$, yielding the specification below in STL-GO
\begin{align}
    \phi_4 \coloneqq \mathbf{G}_{[0, 2]} \bigwedge_{j \in \mathcal{V} \setminus \{i\}}  \pi^{\boldsymbol{\mu}_{d,j}} \to j.(\mathbf{In}^\exists_{\{\mathcal{G}^{s,i}, \mathcal{G}^{c,i}\}, \{1\}} \top), \nonumber
\end{align}   
where $i = 1$, $\boldsymbol{\mu}_{d,j} = 1 - ||x^i - x^j||_2$, and Graphs $\mathcal{G}^{s,i}$ and $\mathcal{G}^{c,i}$ are the sensing and communication graphs only with nodes $i$ and its neighbors, and edges between $i$ and its neighbors as in Example \ref{exm:redundancy2}.
\begin{table}[h]
    \centering
    \renewcommand{\arraystretch}{1.2}
    \begin{tabular}{|l|ccc|ccc|ccc|}
        \hline
        & \multicolumn{3}{c|}{$\sigma = 4$}
        & \multicolumn{3}{c|}{$\sigma = 10$}
        & \multicolumn{3}{c|}{$\sigma = 50$} \\ \cline{2-10}
        & \# sat & \# vio & avg time (ms)
        & \# sat & \# vio & avg time (ms)
        & \# sat & \# vio & avg time (ms) \\ \hline
        $\varphi_{3}$   & 76 & 5 & $1.59 \times 10^{-3}$ & 74 & 7 & $3.08 \times 10^{-3}$ & 59 & 22 & $1.37 \times 10^{-2}$ \\
        $\varphi_{4}$   & 10 & 71 & $2.21 \times 10^{-3}$ & 22 & 59 & $4.42\times 10^{-3}$ & 38 & 43 & $1.70\times 10^{-2}$ \\
        $\phi_{3}$& 76 & 5 & $5.48 \times 10^{-3}$ & 64 & 17 & $2.62 \times 10^{-2}$ & 0 & 81 & $6.07 \times 10^{-1}$ \\
        $\phi_{4}$& 64 & 17 & $8.49 \times 10^{-3}$ & 61 & 20 & $2.04 \times 10^{-2}$ & 33 & 48 & $1.08 \times 10^{-1}$ \\ \hline
    \end{tabular}
    \caption{Results for $\sigma \in \{4, 10, 50\}$ of Section \ref{subsec:drone}, where $\sigma$ refers to number of agents.}
    \label{tab:sigma_small}
\end{table}

\begin{table}[h]
    \centering
    \renewcommand{\arraystretch}{1.2}
    \begin{tabular}{|l|ccc|ccc|}
        \hline
        & \multicolumn{3}{c|}{$\sigma = 100$}
        & \multicolumn{3}{c|}{$\sigma = 500$} \\ \cline{2-7}
        & \# sat & \# vio & avg time (ms)
        & \# sat & \# vio & avg time (ms) \\ \hline
        $\varphi_{3}$ & 64 & 17 & $2.98 \times 10^{-2}$ & 65 & 16 & $1.71 \times 10^{-1}$\\
        $\varphi_{4}$ & 61 & 20 & $3.49 \times 10^{-2}$ & 38 & 43 & $1.72 \times 10^{-1}$\\
        $\phi_{3}$    & 0 & 81 & 2.33 & 0 & 81 & 85.58\\
        $\phi_{4}$    & 17 & 64 & $2.16 \times 10^{-1}$ & 14 & 67 & 1.19\\ \hline
    \end{tabular}
    \caption{Results for $\sigma\!\in\!\{100,500\}$ of Section~\ref{subsec:drone}, where $\sigma$ refers to number of agents.}
    \label{tab:sigma_large}  
\end{table}
Note that formula $\varphi_3$ and $\varphi_4$ do not involve state information and thus any distributed offline monitoring will be equivalent to the centralized offline monitoring results. Therefore, we focus on the centralized monitoring of $\varphi_3$, $\varphi_4$, $\phi_3$, and $\phi_4$. Specifically, we are interested in monitoring the formulas against a window of $t \in \{0, \hdots, 80\}$ (again with unit of minutes), where we monitor for an interval of $[t, t]$ for each $t \in \{0, \hdots, 80\}$ separately. To evaluate the scalability of our algorithm, we present the monitoring results and timing with $\sigma$ in \{4, 10, 50, 100, 500\} in Table \ref{tab:sigma_small} and Table \ref{tab:sigma_large}, where $\#$ sat and $\#$ vio denote the number of satisfactions and violations respectively and avg time denotes the average monitoring time across $t \in \{0, \hdots, 80\}$ in milliseconds. Note that the runtime increases slightly for $\varphi_3$ and $\varphi_4$ moderately for $\phi_4$ and significantly for $\phi_3$. 
This is because that the monitoring algorithm considers all agents in $\mathcal{V}$ when monitoring $\phi_3$.

\section{Conclusion}
In this paper, we introduced Spatio-Temporal Logic with Graph Operators (STL-GO), a novel framework for specifying and verifying complex multi-agent system (MAS) requirements with multiple network topologies. We extended signal temporal logic with graph operators that capture rich interactions among agents through multiple discrete graphs. We introduced “incoming” and “outgoing” graph operators, which can reason over both spatial and temporal properties with various types of graphs. 
We provided a distributed monitoring algorithm that leverages partial information, enabling individual agents to independently monitoring specifications. We demonstrated the expressiveness of STL-GO and our distributed monitoring approach by two case studies.

\section{Acknowledgement}
This work was partially supported by the National Science Foundation through the following grants: CAREER award (SHF-2048094), CNS-1932620, CNS-2039087, FMitF-1837131, CCF-SHF-1932620, IIS-SLES-2417075, funding by Toyota R\&D and Siemens Corporate Research through the USC Center for Autonomy and AI, an Amazon Faculty Research Award, and the Airbus Institute for Engineering Research. This work does not reflect the views or positions of any organization listed.

\bibliographystyle{ACM-Reference-Format}
\bibliography{MASTL}

\end{document}